\documentclass[prd,superscriptaddress,nofootinbib,preprint]{revtex4}
\usepackage{graphicx}
\usepackage{dcolumn}
\usepackage{bm}
\usepackage{amssymb}
\usepackage{mathrsfs}
\usepackage{amsmath}
\usepackage{epsfig}
\usepackage[dvips]{color}
\usepackage{hhline}
\usepackage[colorlinks]{hyperref}
\usepackage[normalem]{ulem}

 \usepackage{hhline} \usepackage[colorlinks]{hyperref}
\usepackage{dcolumn}
\usepackage{bm}
\usepackage{amssymb}
\usepackage{mathrsfs}
\usepackage{multirow}
\usepackage{amsmath}
\usepackage{slashed}
\newcommand{\mathsym}[1]{{}}

\newcommand{\baz}{\begin{array}{cc}}
\newcommand{\bad}{\begin{array}{ccc}}
\newcommand{\ba}{\begin{array}{c}}
\newcommand{\ea}{\end{array}}
\newcommand{\be}{\begin{equation}}
\newcommand{\ee}{\end{equation}}
\newcommand{\bea}{\begin{eqnarray}}
\newcommand{\eea}{\end{eqnarray}}
\newcommand{\nn}{\nonumber}
\newcommand{\bi}{\begin{itemize}}
\newcommand{\ei}{\end{itemize}}
\newcommand{\bmt}{\begin{pmatrix}}
\newcommand{\emt}{\end{pmatrix}}
\newcommand{\bt}{\begin{tabular}}
\newcommand{\et}{\end{tabular}}

\newcommand{\benu}{\begin{enumerate}}
\newcommand{\eenu}{\end{enumerate}}
\begin{document}
\title{Post-Sphaleron baryogenesis and $n-\overline{n}$ oscillation in non-SUSY $SO(10)$ GUT with 
gauge coupling unification and proton decay}
\author{Sudhanwa Patra}
\email{sudha.astro@gmail.com}
 \affiliation{Center of Excellence in Theoretical and Mathematical Sciences, 
Siksha \textquoteleft O\textquoteright Anusandhan University, Bhubaneswar-751030, India}
 \author{Prativa Pritimita}
\email{pratibha.pritimita@gmail.com}
 \affiliation{Center of Excellence in Theoretical and Mathematical Sciences, 
Siksha \textquoteleft O\textquoteright Anusandhan University, Bhubaneswar-751030, India}
\begin{abstract}
Post-sphaleron baryogenesis\textquotedblright  , a fresh and profound mechanism of baryogenesis 
 accounts for the matter-antimatter asymmetry of our present universe in a framework of 
Pati-Salam symmetry. We attempt here to embed this mechanism in a non-SUSY SO(10) grand unified theory by reviving a 
novel symmetry breaking chain with Pati-Salam symmetry as an intermediate symmetry 
breaking step and as well to address post-sphaleron baryogenesis and neutron-antineutron oscillation in a rational manner. 
The Pati-Salam symmetry based on the gauge group $SU(2)_L \times SU(2)_{R} \times SU(4)_C$ is realized in our model 
at $10^{5}-10^{6}$ GeV and the mixing time for the neutron-antineutron oscillation process having $\Delta B=2$ 
is found to be $\tau_{n-\bar{n}} \simeq 10^{8}-10^{10}\,\mbox{secs}$ with the model parameters which is within the reach of 
forthcoming experiments. Other novel features of the model includes low scale right-handed $W^{\pm}_R$, $Z_R$ gauge bosons, 
explanation for neutrino oscillation data via gauged inverse (or extended) seesaw mechanism and most importantly 
TeV scale color sextet scalar particles responsible for observable $n-\bar{n}$ oscillation which can be accessible to 
LHC. We also look after gauge coupling unification and estimation of proton life-time with and without the addition of 
color sextet scalars.  
\end{abstract}
\maketitle
\section{INTRODUCTION}
The Standard Model (SM) of particle physics has given us enough reasons to look beyond its framework for dealing 
with issues like tiny neutrino masses, matter-antimatter asymmetry of the present universe, Dark matter and Dark 
energy, coupling unification of three fundamental interactions. Among all these, the observed baryon asymmetry 
of the universe has motivated the scientific community to work upon it since a long time. The WMAP satellite data 
\cite{Dunkley:2008,Komatsu:2011}, when combined with large scale structures (LSS) data, gives the baryon asymmetry 
of the universe to be $\eta^{\mbox{\small CMB}} \simeq (6.3\pm 0.3) \times 10^{-10} $ while an independent measurement 
of baryon asymmetry carried out by BBN \cite{Yao:2006} yields $\eta^{\mbox{\small BBN}} \simeq (3.4 -6.9) \times 10^{-10}$. 
Two compelling mechanisms namely Leptogenesis \cite{Fukugita:1986} and  Weak scale baryogenesis \cite{EWbary} have been 
prime tools for explaining baryon asymmetry of the universe. In leptogenesis the desired lepton asymmetry is created 
by the lepton number violating as well as out of equilibrium decays of heavy particles which is subsequently converted 
into baryon asymmetry by the non-perturbative ($B+L$)-violating sphaleron interactions \cite{Buchmuller:2005eh,Davidson:2008bu}. 

An inadequate knowledge about the nature of new physics beyond the standard model 
leaves us with no choice but to explore all possibilities which may explain the origin of matter-antimatter asymmetry. 
Recently a new idea behind baryon asymmetry has been explored named "Post-Sphaleron baryogenesis (PSB)" which 
occurs via the decay of a scalar boson singlet under standard model having mass around few hundreds of GeV and 
a high dimensional baryon number violating coupling \cite{Babu:2006xc, Babu:2012vc,Babu:2013yca}, where the 
Yukawa coupling(s) of the scalar(s) act as the 
source of CP-asymmetry. Apparently,this high dimensional baryon number violating coupling is generated via new physics 
operative beyond standard model electroweak theory. The mechanism of PSB is based on the idea that 
the required amount of baryon asymmetry of the universe can be generated below the scale of electroweak phase transition 
where the sphaleron has decoupled from the Hubble expansion rate. Although the proposal seems interesting it has 
not yet been incorporated in a realistic grand unified theory. Hence we attempt here to embed the proposal of PSB 
in a non-SUSY SO(10) GUT with Pati-Salam (PS) symmetry and Left-Right (LR) symmetry as intermediate symmetry breaking steps.

A detail study of the literatures \cite{Mohapatra:1974gc,Pati:1974yy,Senjanovic:1975rk,
Mohapatra:1980yp,Mohapatra:1979ia,Deshpande:1990ip,Lazarides:1980nt,Dev:2013oxa} 
gives an idea about many intriguing features of the $SO(10)$ 
grand unified theory (including both non-SUSY and SUSY). One of these features is that when left-right gauge symmetry 
appears as an intermediate symmetry breaking step in a novel symmetry breaking chain, then seesaw mechanism can be naturally 
incorporated into it. In conventional seesaw models associated with thermal leptogenesis the mass scale for heavy 
RH Majorana neutrino is at $10^{10}$ GeV which makes it unsuitable for direct detectability at current 
accelerator experiments like LHC. Therefore, it is necessary to construct a theory having 
$SU(2)_L \times SU(2)_{R} \times U(1)_{B-L} \times SU(3)_C$ and $SU(2)_L \times SU(2)_{R} \times SU(4)_C$ 
gauge groups as intermediate symmetry breaking steps which results in low mass right-handed Majorana neutrinos along with 
$W_R$, $Z^\prime$ gauge bosons at TeV scale. At the same time it should be capable of explaining post-sphaleron 
baryogenesis elegantly along with other derivable predictions like proton decay and neutron-antineutron oscillation.

We intend to discuss TeV scale post-sphaleron baryogenesis, neutron-antineutron oscillation having mixing time close 
to the experimental limit with the Pati-Salam symmetry or $SO(10)$ GUT as mentioned in a recent work \cite{Awasthi:2013ff} 
slightly modifying the Higgs content where non-zero light neutrino masses can be accommodated via gauged extended 
inverse seesaw mechanism along with TeV scale $W_R$, $Z^\prime$ gauge bosons. As discussed in the work \cite{Awasthi:2013ff} 
the Dirac neutrino mass matrix is similar to the up-quark mass matrix even with low scale right-handed symmetry breaking. 
Though the details has been already discussed in the above mentioned work we breifly clarify the point as follows.



In non-SUSY $SO(10)$, the type I seesaw \cite{typeI} contribution to neutrino mass is given by 
$$m^{I}_\nu = - M_D M^{-1}_R M^T_D\,, $$
where $M_D$ is the Dirac neutrino mass matrix, $M_R$ is the Majorana neutrino mass matrix for right-handed 
neutrinos and is related to the right-handed symmetry breaking scale. The Dirac neutrino mass matrix 
and up-quark mass matrices are similar in a generic $SO(10)$ model that has high scale Pati-Salam symmetry 
as an intermediate breaking step relating quarks and leptons with each other. Hence, $M_D \simeq M_u$, which further 
implies that the $\tau-$ neutrino Dirac Yukawa coupling should be equal to top-quark Yukawa coupling. 
With $M_D\simeq M_u \simeq 100$ GeV, the sub-eV scale of light neutrino consistent with oscillation data 
requires the right-handed scale (seesaw scale) to be greater than $10^{13}$ GeV. Such high seesaw scale 
makes this idea difficult to be probed at any foreseeable laboratory experiments. 
Hence, as an alternative way, emphasizing on its verifiability at LHC, inverse seesaw 
mechanism \cite{inv,Bdev-non} has been proposed, with an extra $SO(10)$ fermion singlet $S$ (in addition to the 
existing fermion content of $SO(10)$), with light neutrino mass formula 
$$m_\nu =\left(\frac{M_D}{M}\right) \mu \left(\frac{M_D}{M}\right)^T\, ,$$
where $M$ is the $N-S$ mixing matrix and $\mu$ is the small lepton number violating mass term for 
sterile neutrino $S$. The above relation can be recasted as
$$\left( \frac{m_\nu}{\mbox{0.1\, eV}}\right) = \left(\frac{M_D}{\mbox{100\, GeV}} \right)^2 
  \left(\frac{\mu}{\mbox{keV}}\right) \left(\frac{M}{10^4\, \mbox{GeV}} \right)^{-2}\,.$$
Hence, sub-eV mass for light neutrinos are consistent with $M_D \simeq M_u$ (or, $Y_D \simeq Y_t$) which 
is a generic predictions of high scale Pati-Salam symmetry and compatible with low right-handed 
symmetry breaking scale ($M_R$) since inverse seesaw formula is independent of $M_R$.
We have utilised this particular property of low scale right-handed symmetry breaking in studying 
Post-sphaleron baryogenesis and neutron-antineutron oscillation even though a complete discussion 
on the origin of neutrino masses and mixing via low sacle extended inverse seesaw has been omitted.

Here we sketch out the complete work of our paper. In Sec.II, we briefly discuss 
non-SUSY $SO(10)$ GUT with a novel symmetry breaking chain, having $\mathcal{G}_{2213}$ and $\mathcal{G}_{224}$ 
as intermediate symmetry breaking steps. In Sec.III we show how gauge coupling unification is achieved in our model. 
In Sec.IV we discuss the TeV scale post-sphaleron baryogenesis and embed it within the novel chain of non-SUSY $SO(10)$ 
model with the self-consistent model parameters. In Sec.V, we estimate the mixing 
time for neutron-antineutron oscillation. In Sec.VI, we present an idea how low mass scales for RH Majorana neutrino 
as well as right-handed gauge bosons $W_R$, $Z^\prime$ are allowed in the model, while explaining light neutrino masses 
via gauged extended seesaw mechanism. In Sec.VII we conclude our work with results and summary including a note on 
viability of the model at LHC.

\section{THE MODEL}
In this section we shall discuss the one-loop gauge coupling unification and estimate the proton life time including 
short distance enhancement factor to the $d=6$ proton decay operator by reviving the symmetry breaking chain \cite{Awasthi:2013ff}
{\small 
\begin{eqnarray}
SO(10)
& &\stackrel{M_U}{\longrightarrow}SU(2)_L \times SU(2)_{R} \times SU(4)_C \times D \quad 
                           \left[\mathcal{G}_{224D}, \, \, (g_{2L} = g_{2R})  \right]\nonumber \\ 
& &\hspace*{-1.5cm} \mathop{\longrightarrow}^{M_P}_{} SU(2)_L \times SU(2)_{R} \times SU(4)_C 
              \quad \left[\mathcal{G}_{224}, \, \, (g_{2L} \neq g_{2R})  \right]\nonumber \\ 
& &\hspace*{-1.5cm} \mathop{\longrightarrow}^{M_C}_{} SU(2)_L \times SU(2)_{R} \times U(1)_{B-L} \times SU(3)_C 
              \quad \left[\mathcal{G}_{2213} \, \right]\nonumber \\ 
& &\hspace*{-1.5cm} \mathop{\longrightarrow}^{M_\Omega}_{} SU(2)_L \times U(1)_{R} \times U(1)_{B-L} \times SU(3)_C 
              \quad \left[\mathcal{G}_{2113} \, \right]\nonumber \\ 
& &\hspace*{-1.5cm} \mathop{\longrightarrow}^{M_{B-L}}_{} SU(2)_L \times U(1)_{Y} \times SU(3)_C 
              \quad \left[\mathcal{G}_{\rm SM} \equiv \mathcal{G}_{\rm 213}\right] \nonumber \\
& &\hspace*{-1.5cm} \mathop{\longrightarrow}^{M_Z}_{}~U(1)_{\rm em}\times SU(3)_C \quad \quad   \left[\mathcal{G}_{\rm 13}\right]\, .
\label{chain}              
\end{eqnarray}
}
The chain breaks in a sequence, where $SO(10)$ first breaks down to $\mathcal{G}_{224D}, \, \, (g_{2L} = g_{2R})$ after 
the Higgs representation $\langle (1,1,1) \rangle \subset \{54\}_H$ is given a VEV, then the spontaneous breakdown of 
D-parity occurs in $\mathcal{G}_{224D}, \, \, (g_{2L} = g_{2R}) \to \mathcal{G}_{224}, \, (g_{2L} \neq g_{2R})$ with the 
assignment of VEV to D-parity odd component $\langle(1,1,1) \rangle$ contained in the Higgs representation 
$\{210\}_H$. The decomposition of $\{210\}_H$ under $\mathcal{G}_{224}$ is
\begin{eqnarray}
\{210\}_H&=&(1,1,1) \oplus (2,2,20) \oplus (3,1,15) \oplus (1,3,15) \nonumber \\
& &\quad \oplus (2,2,6) \oplus (1,1,15)\, .
\end{eqnarray}
Spontaneous D-parity mechanism is aptly utilized here, since the theory allows low mass scale for right-handed 
Higgs fields around $\mathcal{O}$(TeV) while keeping all its left-handed components at D-parity breaking scale. Now assigning 
a VEV to the neutral component $\langle(1,1,15)\rangle \subset \{210\}_H$, the Pati-Salam symmetry ($\mathcal{G}_{224}$) 
breaks down to left-right symmetry ($\mathcal{G}_{2213}$). The next step of symmetry breaking $\mathcal{G}_{2213} \to 
\mathcal{G}_{2113}$ occurs via the VEV $\langle(1,3,0,1)\rangle \subset \{210\}_H$. The right-handed gauge 
boson $W_R$ acquires a mass in the range of few TeV and contributes sub-dominantly to neutrinoless 
double beta decay. 

The most desirable symmetry breaking step $\mathcal{G}_{2113} \to \mathcal{G}_{213}$ is achieved by the $\{126\}_H$ of $SO(10)$ 
though we have added another Higgs representation $\{16\}_H$ for realization of gauged inverse seesaw mechanism operative 
at TeV scale. The decomposition of the Higgs $\{126\}_H$ under $\mathcal{G}_{224}$ is
\begin{eqnarray}
\{126\}_H&=&(3,1,10) \oplus (1,3,\overline{10}) \oplus (2,2,15) \oplus (1,1,6) \nonumber \\
\end{eqnarray}
As we have pointed earlier, due to D-parity mechanism, the right-handed triplet Higgs field 
$\Delta_R (1,3,-2,1)$ contained in $(3,1,10)$ gets its mass at TeV scale while its left-handed partner 
$\Delta_L (3,1,-2,1)$ has its mass at D-parity breaking scale $M_P$. As a result of this symmetry breaking, 
the neutral component of right-handed gauge boson $Z^\prime$ gets its mass around $\mathcal{O}$(TeV) with 
the experimental bound $M_{W_R} \geq 2.5$~TeV \cite{CMS:2012zv,ATLAS:2012ak}. The final stage of symmetry breaking $\mathcal{G}_{2113} \to \mathcal{G}_{213}$ 
is carried out by giving VEV to the neutral component of SM Higgs doublet $\langle \phi^0(2,1/2,1) \rangle$ 
contained in the bidoublet $\Phi \subset \{10\}_H$.

We shall now check whether $SO(10)$ having TeV scale post-sphaleron baryogenesis, neutron-antineutron oscillation 
and gauged inverse seesaw mechanism is consistent with gauge coupling unification. It is found that the coupling constants 
unify at ($10^{17}-10^{18.5}$) GeV with the Higgs fields {\bf $\{10\}_H$+$\{10\}_H^\prime$ + $\{16\}_H$ + $\{126\}_H$ 
+$\{210\}_H$}. Some good reasons behind taking these Higgs fields are; firstly, the TeV scale post-sphaleron baryogenesis 
and neutron-antineutron oscillation can be well explained with these parameters while predicting $W_R$ gauge boson 
in TeV range; secondly, it allows $B-L$ breaking ($M_{B-L}$) at TeV scale resulting 
$Z^\prime$ mass $\geq 1.6$ TeV, moreover it explains tiny masses for light neutrinos consistent 
with neutrino oscillation data via TeV scale gauged inverse seesaw mechanism and LFV decays with branching ratios 
accessible to ongoing search experiments.

\section{GAUGE COUPLING UNIFICATION {\small AND} PROTON DECAY}
\label{sec:rge}
\subsection{One-loop renormalization group equations (RGEs) for gauge coupling evolution}
\label{subsec1:rge}
For simplicity, we consider only the one-loop renormalization group equations(RGEs) for gauge coupling evolution 
which can be written as
\begin{eqnarray}
\mu\,\frac{d\,g_{i}}{d\,\mu}=\frac{a_i}{16 \pi^2} g^{3}_{i}\, \quad {\large \Longrightarrow}\, \quad
\frac{d\, \alpha^{-1}_{i}}{d\,t}=\frac{\pmb{a_i}}{2 \pi}
\end{eqnarray}
where, $t=\ln(\mu)$, $\alpha_{i}=g^2_{i}/(4 \pi)$ is the fine structure constant, and $\pmb{a_i}$ is the one-loop beta coefficients derived for the the corresponding 
$i^{\rm th}$ gauge group for which coupling evolution has to be determined. Using the input parameters, 
electroweak mixing angle $\sin^2\theta_W(M_Z)=0.2312$, electromagnetic coupling constant $\alpha(M_Z) = 127.9$ 
and strong coupling constant $\alpha_{S}(M_Z) = 0.1187$ taken from PDG \cite{Yao:2006,pdg} 
the values of three coupling constants at electroweak scale $M_Z=91.187$ GeV can be calculated precisely to be,  
\begin{equation}
\left(\begin{array}{cc}
\ \alpha_{2L}(M_Z) \\
\ \alpha_{1Y}(M_Z) \\
\ \alpha_{3C} (M_Z) 
\end{array}\right)
= \left(\begin{array}{cc}
\ 0.033493^{+0.000042}_{-0.000038} \\
\ 0.016829 \pm 0.000017 \\
\ 0.118 \pm 0.003 
\end{array}\right)\, ,
\label{pdg:alphas}
\end{equation}
where $\{\alpha_{2L}(M_Z), \alpha_{1Y}(M_Z), \alpha_{3C}(M_Z) \}$ denote the fine structure constants for the 
SM gauge group $\mathcal{G}_{213}=SU(2)_L \times U(1)_Y \times SU(3)_C$.

\subsection{Higgs content for the model and corresponding one-loop beta coefficients $\pmb{a_i}$}
\label{subsec2:rge}
The Higgs contents for the model used in different ranges of mass scales under respective gauge symmetries ($\mathcal{G}_I$) 
with a particular symmetry breaking chain as considered in a recent work \cite{Awasthi:2013ff} where the prime interest was to keep 
the $W_R$, $Z_R$ gauge bosons at TeV scale are as follows,
\begin{eqnarray}
& &\hspace*{-1.0cm} {\bf \mbox{(i)}\, \mu=M_Z - M_{B-L} }:  G={\rm SM} = G_{213}, 
             \hspace*{0.2cm} \mbox{Higgs:\,}\Phi (2,1/2,1)\, ; \nonumber \\
& &\hspace*{-1.0cm} {\bf \mbox{(ii)}\, \mu=M_{B-L} - M_\Omega}: G= G_{2113}, \nonumber \\
& & \hspace*{0.0cm} \mbox{Higgs:\,} \Phi_1 (2,1/2,0,1) \oplus \Phi_2 (2,-1/2,0,1) \oplus \chi_R (1,1/2,-1,1) \oplus \Delta_R (1,1,-2,1) \, ; \nonumber \\ 
& &\hspace*{-1.0cm} {\bf \mbox{(iii)}\, \mu=M_\Omega - M_C}: G= G_{2213},
\nonumber \\
& & \hspace*{0.0cm} \mbox{Higgs:\,}\Phi_1 (2,2,0,1) \oplus \Phi_2 (2,2,0,1) \oplus 
\chi_R (1,2,-1,1)\oplus \Delta_R (1,3,-2,1)\oplus \Omega_R(1,3,0,1) \nonumber \\
\label{higgs-a}
\end{eqnarray}
\begin{eqnarray}
& &\hspace*{-1.0cm} {\bf \mbox{(iv)}\, \mu=M_C - M_{\xi}}: G= G_{224}, \nonumber \\
& &\hspace*{0.2cm} \mbox{Higgs:\,}\Phi_1(2,2,1)_{10}\oplus \Phi_2(2,2,1)_{10^\prime} \oplus \Delta_R(1,3,\overline{10})_{126} 
                    \oplus \chi_R(1,2,\overline{4})_{16} \nonumber \\
& & \hspace*{1.2cm}\oplus \Omega_R(1,3,15)_{210} \oplus \Sigma(1,1,15)_{210} \nonumber \\
& &\hspace*{-1.0cm} {\bf \mbox{(v)}\, \mu=M_\xi - M_{P}}: G= G^\prime_{224}, \nonumber \\
& &\hspace*{0.2cm} \mbox{Higgs:\,}\Phi_1(2,2,1)_{10}\oplus \Phi_2(2,2,1)_{10^\prime} \oplus \Delta_R(1,3,\overline{10})_{126} 
                    \oplus \chi_R(1,2,\overline{4})_{16} \nonumber \\
& & \hspace*{1.2cm}\oplus \Omega_R(1,3,15)_{210} \oplus \Sigma(1,1,15)_{210} \oplus \xi(2,2,15)_{126} \nonumber \\
& &\hspace*{-1.0cm} {\bf \mbox{(vi)}\, \mu=M_P - M_U}: G= G_{224D}, \nonumber \\
& &\hspace*{0.2cm} \mbox{Higgs:\,}
\Phi_1(2,2,1)_{10}\oplus \Phi_2(2,2,1)_{10^\prime} \oplus \Delta_L(3,1,10)_{126} \oplus \Delta_R(1,3,\overline{10})_{126} \nonumber \\
& &\hspace*{1.2cm} \oplus \chi_L(2,1,4)_{16} \oplus \chi_R(1,2,\overline{4})_{16} \oplus \Omega_L(3,1,15)_{210} \oplus \Omega_R(1,3,15)_{210}\nonumber \\
& &\hspace*{1.2cm}\oplus \Sigma(1,1,15)_{210} \oplus \xi(2,2,15)_{126} \oplus \sigma(1,1,1)_{210}\, .\nonumber \\
\label{higgs-b}
\end{eqnarray} 
Here we find two categories of Higgs spectrum; {\bf Model-I\,} having Higgs spectrum as given in eqn.(\,\ref{higgs-a}) 
and eqn.(\,\ref{higgs-b}) excluding the bitriplet Higgs scalar which estimates a proton life time that is far from the reach of 
search experiments and {\bf Model-II\,} having the same Higgs spectrum, including the bitriplet Higgs scalar 
$(3,3,1) \subset \mathcal{G}_{224}$ from mass scale $M_C$ onwards which estimates a proton life time very close to the 
experimental limit. Thus {\bf Model-II\,} serves our purpose. 

The one-loop beta coefficients are found to be the same for both the models at mass scale ranges $M_Z - M_{B-L}$, 
$M_{B-L}-M_\Omega$, and $M_\Omega - M_{C}$ i.e., 
\begin{eqnarray}
& &\hspace*{-1.0cm} {\bf \mbox{(i)}\, \mu=M_Z - M_{B-L} }:  G={\rm SM} = G_{2_L 1_Y 3_C}, \quad \pmb{a_i}=\left(-19/6,\, 41/10,\, -7\right)
\nonumber \\ 
& &\hspace*{-1.0cm} {\bf \mbox{(ii)}\, \mu=M_{B-L} - M_\Omega}: G= G_{2_L 1_{R} 1_{(B-L)} 3_C}, 
                                       \quad \pmb{a_i}=\left(-3,\, 19/4,\, 37/8,\, -7\right)
                                       \nonumber \\ 
& &\hspace*{-1.0cm} {\bf \mbox{(iii)}\, \mu=M_\Omega - M_C}: G= G_{2_L 2_{R} 1_{(B-L)} 3_C}, 
                                       \quad \pmb{a_i}=\left(-8/3,\, -2/3,\, 23/4,\, -7\right)\, ,
\end{eqnarray}
whereas, they differ at Pati-Salam scale $M_C$ to the Unification scale $M_U$ as shown in Table.\ref{tab:beta_coeff}. 

\begin{table*}[htb]
\centering
\begin{tabular}{|c|c|c|c|}
\hline
&&& \\[-4mm]
$G_{I}$  & {\bf Mass ranges}  & $\pmb{a_i}$ for {\bf Model-I}  & $\pmb{a_i}$ for {\bf Model-II}   \\[4mm]
\hline \hline
&&&\\[-4mm]
${\small G_{2_L2_R4_C}}$ 
& {\bf $M_C-M_{\xi}$}
&
$\bmt -8/3 \\
       29/3\\
      -14/3 \emt$
&
$\bmt -2/3 \\
       35/2\\
      -14/3 \emt$\\[7mm]
\hline 
&&&\\[-4mm]
${\small G_{2_L2_R4_C}}$ 
&{\bf $M_{\xi}-M_P$}
&
$\bmt 7/3 \\
      44/3\\
      2/3 \emt$
&
$\bmt -12/3 \\
      35/3\\
      -14/3 \emt$\\[7mm]
\hline 
&&&\\[-4mm]
${\small G_{2_L2_R4_C D}}$ 
&{\bf $M_P-M_{U}$}
&
$\bmt 44/3 \\
      44/3 \\
      6 \emt$                                                  &
$\bmt 35/3 \\
      35/3 \\
      2/3 \emt$\\[7mm]
\hline
\end{tabular}
\caption{One-loop beta coefficients for different gauge coupling evolutions, without Bitriplet Higgs scalar in {\bf Model-I} and with 
         a Bitriplet Higgs scalar (3,3,1) under the Pati-Salam group $SU(2)_L \times SU(2)_R \times SU(4)_C$ in {\bf Model-II}.}
\label{tab:beta_coeff}
\end{table*}
 
\begin{figure*}[htb!]
\begin{minipage}[t]{0.49\textwidth}
\hspace{-0.4cm}
\begin{center}
\includegraphics[scale=0.85,angle=0]{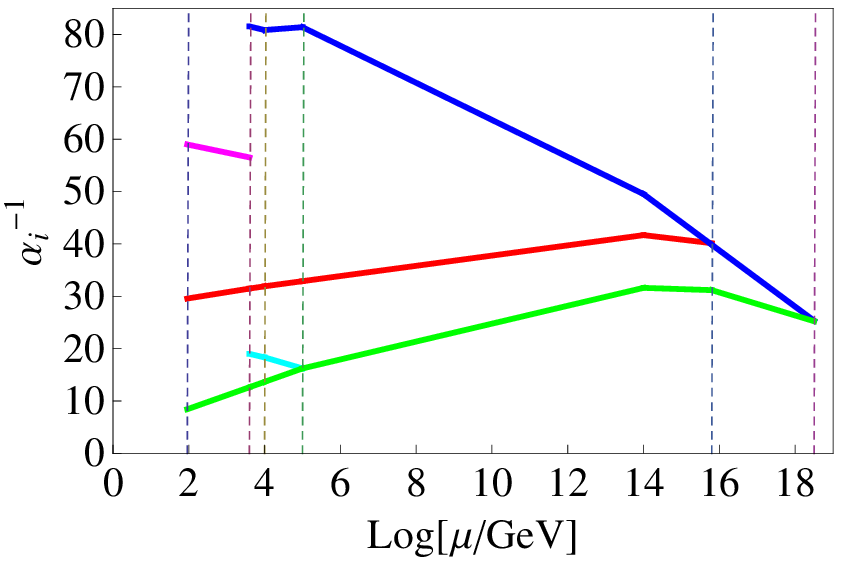}
\end{center}
 \end{minipage}
 \hfill
 \begin{minipage}[t]{0.49\textwidth}
 \hspace{-0.4cm}
 \begin{center}
 \includegraphics[scale=0.85,angle=0]{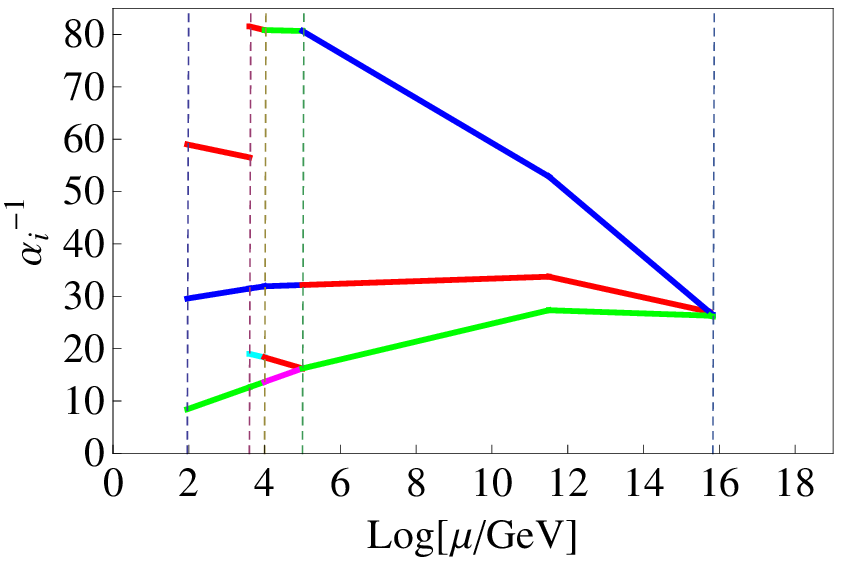}
 \end{center}
 \end{minipage}
 \caption{Gauge coupling evolution plot having TeV scale $W_R$, $Z_R$ bosons where $M_U=2.65\times 10^{15.8}$ GeV}
\label{lightneutrino_general}
 \end{figure*}
The gauge coupling unification for this work is shown in Fig.\,\ref{lightneutrino_general} 
with the allowed mass scales desirable for our model predictions,
\bea 
& &M_{B-L}= 4-7~{\rm TeV},\, M_{\Omega} = 10~{\rm TeV},\, M_C = 10^{5}-10^{6}\, \mbox{GeV}\, ,\nn \\
& &M_P\simeq 10^{15.65}~{\rm GeV\,\, and\,\,} M_{\rm G}\simeq 10^{18.65}~{\rm GeV} \, .
\eea

\subsection{Estimation of Proton life time for $p \to \pi^0\, e^+$}
\label{subsec3:proton}
The decay rate for the gauge boson mediated proton decay in the channel $p \to \pi^0\, e^+$ including strong and electroweak 
renormalization effects on the ${\rm d}=6$ operator starting from the GUT scale to the proton mass (i.e, 1 GeV) \cite{Babu:1992ia, Bertolini:2013vta} 
comes out to be 
\begin{eqnarray}  
\Gamma\left( p\rightarrow \pi^0 e^+ \right) &=&\frac{\pi}{4}\, A^2_L\, \frac{|\overline{\alpha}_H|^2}{f^2_\pi} 
 \frac{m_p\, \alpha^2_U}{M^4_U}  \left(1 + \mathcal{F} + \mathcal{D} \right)^2 \mathcal{R}\, .
\label{decay-width-proton}
\end{eqnarray}
In the eq.\,(\ref{decay-width-proton}), $A_L=1.25$ is renormalization factor from the electroweak scale to the proton mass, $\mathcal{D}=0.81$, 
$\mathcal{F}=0.44$, $\overline{\alpha}_H=-0.011\, \, \mbox{GeV}^3$, and $f_\pi=139\, \, \mbox{MeV}$ which have 
been extracted as phenomenological parameters by the chiral perturbation theory and lattice gauge theory. 
Also $m_p=938.3\, \, \mbox{MeV}$ is the proton mass, and $\alpha_U \equiv \alpha_G$ is the gauge fine structure constant derived 
at the GUT scale. It is worth to note here that the renormalization factor $\mathcal{R}=\left[\left(A_{SR}^2+A_{SL}^2\right)
\left(1+ |{V_{ud}}|^2\right)^2\right]$ for $SO(10)$, $V_{ud}=0.974=$  with $A_{SL}(A_{SR})$ being the short-distance 
renormalization factor in the left (right) sectors, and $V_{ud}$ is the  $(1,1)$ element of $V_{CKM}$ for quark mixings. 

After re-expressing $\alpha_H = \overline{\alpha}_H \left(1 + \mathcal{F} + \mathcal{D} \right) = 0.012\, \, \mbox{GeV}^3$, and 
$\mathcal{A}_R\simeq \mathcal{A}_{L} \mathcal{A}_{SL} \simeq \mathcal{A}_{L} \mathcal{A}_{SR}$, the proton life time can 
be expressed as 
\begin{eqnarray}  
\tau_p = \Gamma^{-1}\left( p\rightarrow \pi^0 e^+ \right) &=&\frac{4}{\pi}\,\frac{f^2_\pi}{m_p}\frac{M^4_U}{\alpha^2_U} 
         \frac{1}{\alpha^2_H \mathcal{A}^2_{R}} \frac{1}{\mathcal{F}_q} \, ,
\label{lifetime-proton}
\end{eqnarray}
where $\mathcal{F}_q \simeq 7.6$.

\noindent
{\bf Short distance enhancement factor $\mathcal{A}_{SL}$ extrapolated from GUT scale to $1$ GeV:}
For estimating proton decay rate in the channel $p \to e^+\pi^0$ having dimension-6 operator, one needs to extrapolate 
the operator from the GUT scale physics to the low energy physics at the scale of $m_p=1\, \mbox{GeV}$ \cite{Ibanez:1984ni, Buras:1977yy, 
BhupalDev:2010he}. With the particular symmetry breaking chain allowed in the non-SUSY $SO(10)$ model (following the ref.\, \cite{BhupalDev:2010he}), 
the whole energy range can be separated into following parts
\begin{enumerate}
\item[${\bf \tiny I.}$]   from non-SUSY $S0(10)$ GUT scale,$M_U$, to the Pati-Salam symmetry with D-parity ($\mathcal{G}_{224D}$, 
                          $g_{2L}=g_{2R}$) invariance scale, $M_P$\, ,
\item[${\bf \tiny II.}$]  from $M_P$ to the Pati-Salam symmetry without D-parity ($\mathcal{G}_{224}$, $g_{2L} \neq g_{2R}$) 
                          scale $M_C$\, ,
\item[${\bf \tiny III.}$] from $M_C$ to $SU(4)_C$ breaking scale, $M_\Omega$, where we have left-right symmetric model 
                          ({\bf LRSM}) $\mathcal{G}_{2213}$\, ,  
\item[${\bf \tiny IV.}$]  from left-right symmetry breaking scale ($M_\Omega$) to $\mathcal{G}_{2113}$ scale ($M_{B-L}$)\, 
\item[${\bf \tiny V.}$]   from $\mathcal{G}_{2113}$ scale ($M_{B-L}$) to standard model $\mathcal{G}_{213}$\, ,
\item[${\bf \tiny V.}$]   from standard model to $1\, \mbox{GeV}$\,. 
\end{enumerate}
As discussed in refs.\, \cite{Ibanez:1984ni, Buras:1977yy, BhupalDev:2010he}, the enhancement factor below SM 
for the $LLLL$ operator is 
$$\mathcal{A}^{\prime}_{L} = \bigg[\frac{\alpha_s (\mbox{1 GeV})}{\alpha_s (m_t)}\bigg]
 ^{-\frac{4}{2 \cdot \left(-11+\frac{2}{3} \, n_f \right)}}\, ,$$
where, $n_f$ denotes the number of quark flavors at the particular energy scale of our interest. 
Neglecting the effect due to $\alpha_{2L}$ and $\alpha_{Y}$ since their contributions are suppressed 
as compared to the strong coupling effect $\alpha_s$, this enhancement factor can be expressed 
in a more explicit manner as
\begin{eqnarray}
      \mathcal{A}^{\prime}_{L}=\bigg[\frac{\alpha_s (\mbox{1 GeV})}{\alpha_s (m_c)}\bigg]^{2/9} 
                               \bigg[\frac{\alpha_s (m_c)}{\alpha_s (m_b)}\bigg]^{6/25} 
                               \bigg[\frac{\alpha_s (m_b)}{\alpha_s (m_t)}\bigg]^{6/23}\, .
\end{eqnarray}
Since the model considered here is non-supersymmetric version of $SO(10)$ GUT, all other enhancement factors 
can be written in the same way as
\begin{eqnarray}
& &\mathcal{A}^{\rm SM}_{SL}=\bigg[\frac{\alpha_i (m_t)}{\alpha_i (M^0_R)}\bigg]
           ^{\frac{- \gamma_i}{2\, {\Large \pmb a_i}}}\, , 
\end{eqnarray}
with $\gamma_i$ (${\bf \large \pmb a_i}$) as the anomalous dimension (one-loop beta coefficients) 
for the corresponding gauge group $i=SU(2)_{L},\, U(1)_{Y},\, SU(3)_{C}$. Similarly, one can write 
the enhancement factor valid for $\mathcal{G}_{2113}$, $\mathcal{G}_{2213}$, $\mathcal{G}_{224}$, 
and $\mathcal{G}_{224D}$ as
\begin{eqnarray}
& &\mathcal{A}^{2113}_{SL}=\bigg[\frac{\alpha_i (M^0_R)}{\alpha_i (M^+_R)}\bigg]
           ^{\frac{- \gamma_i}{2\, {\Large \pmb a_i}}}\, , \mbox{with}\, i=SU(2)_{L},\,U(1)_R,\, U(1)_{B-L},\, SU(3)_{C}\, ,
         \nonumber \\
& &\mathcal{A}^{2213}_{SL}=\bigg[\frac{\alpha_i (M^+_R)}{\alpha_i (M_C)}\bigg]
           ^{\frac{- \gamma_i}{2\, {\Large \pmb a_i}}}\, , \mbox{with}\, i=SU(2)_{L},\,SU(2)_R,\, U(1)_{B-L},\, SU(3)_{C}\, ,
         \nonumber \\
& &\mathcal{A}^{224}_{SL}=\bigg[\frac{\alpha_i (M_C)}{\alpha_i (M_P)}\bigg]
           ^{\frac{- \gamma_i}{2\, {\Large \pmb a_i}}}\, , \mbox{with}\, i=SU(2)_{L},\,SU(2)_R,\, SU(4)_{C}\, ,
         \nonumber \\
& &\mathcal{A}^{224D}_{SL}=\bigg[\frac{\alpha_i (M_P)}{\alpha_i (M_U)}\bigg]
           ^{\frac{- \gamma_i}{2\, {\Large \pmb a_i}}}\, , \mbox{with}\, i=SU(2)_{L},\,SU(2)_R,\, SU(4)_{C}\, \mbox{with D-parity}\, .
         \nonumber
\end{eqnarray}
Hence, the complete short distance enhancement renormalization factor for this $d=6$ proton decay operator is found to be 
\begin{equation}
\mathcal{A}_{SL}=\mathcal{A}^{\rm SM}_{SL} \cdot \mathcal{A}^{2113}_{SL} \cdot \mathcal{A}^{2213}_{SL} 
                 \cdot \mathcal{A}^{224}_{SL} \cdot \mathcal{A}^{224D}_{SL} \,.
\end{equation}

We have earnestly followed the prescription given in ref.\cite{Ibanez:1984ni, Buras:1977yy} for the derivation of anomalous dimension for the 
effective $d=6 (LLLL)$ proton decay operator. With a choice of TeV scale particle spectrum used in our model, the unification 
scale is found to be $M_U=2.65\times 10^{18.5}$ GeV for {\bf Model-I} and $M_U=10^{15.8}$ GeV for {\bf Model-II}. We have estimated 
the factor $\mathcal{A}_{R}=\mathcal{A}_L \cdot \mathcal{A}_{SL}$, approximately, to be $4.36$ with the value of long distance 
renormalization factor $A_L=1.25$ which is the same for both the models. 

With these input parameters, the model under consideration predicts the proton life time to be $$\tau (p \to e^+ \pi^0) = 2.6 
\times 10^{34}\, \mbox{yrs} $$ that is closer to the latest Super-Kamiokande experimental bound \cite{Nishino:2012ipa,babuetal}
\begin{eqnarray}\tau (p \to e^+ \pi^0) \big|_{SK, 2011} > 8.2 \times 10^{33}\, 
\mbox{yrs}\, ,
\end{eqnarray}
and ably supports planned experiments that can reach a bound \cite{hyperk}
\begin{eqnarray}
& &\tau (p \to e^+ \pi^0) \big|_{HK, 2025} > 9.0 \times 10^{34}\, \mbox{yrs} \nonumber \\
& &\tau (p \to e^+ \pi^0) \big|_{HK, 2040} > 2.0 \times 10^{35}\, \mbox{yrs} \nonumber \\
\end{eqnarray}

\section{TEV SCALE POST-SPHALERON BARYOGENESIS}
\label{sec:psb}
\subsection{Basic interaction terms}
\label{subsec1:psb}
As already discussed in Sec.\,\ref{sec:rge}, Pati-Salam symmetry survives till few $100$ TeV scale playing 
an important role in the explanation of baryogenesis mechanism and neutron-antineutron oscillation. We need to know 
all the basic interactions using quarks and di-quarks under high scale Pati-Salam symmetry as well as under 
low scale SM like interactions around TeV scale in order to explain the above said phenomena successfully. For that, 
we take a look at the decomposition of the Pati-Salam Higgs representation 
$\Delta_R(1,3, \overline{10})$ under left-right symmetry group $SU(2)_L \times SU(2)_R \times U(1)_{B-L} \times 
SU(3)_C$ and the SM gauge group $SU(2)_L \times U(1)_{Y} \times SU(3)_C$ 
\begin{eqnarray}
\Delta(1, 3, \overline{10}) &=& \{ \Delta_{\ell \ell} (1, 3, -2, 1)  \oplus \Delta_{q \ell} (1, 3, -2/3, 3^*) 
                               \oplus \Delta_{qq} (1, 3, 2/3, 6^*)  \nonumber \\ & & 
                               \hspace*{6cm}\mbox{under}\, \,\mathcal{G}_{2_L 2_R 1_{B-L} 3_C}\, ,  \\
                               &\supset& \Delta_{\nu \nu} (1,0,1) \oplus \Delta_{\nu e} (1,1,1) \oplus \Delta_{ee} (1,2,1) 
                               \nonumber \\ &\oplus& 
                               \Delta_{u \nu} (1,-2/3,3^*) \oplus \Delta_{d e} (1,1/3,3^*) \oplus \Delta_{u e} (1,1/3,3^*) 
                               \oplus \Delta_{d \nu} (1,1/3,3^*)  \nonumber \\ 
                               &\oplus& \Delta_{u u} (1, -4/3,6^*) \oplus \Delta_{u d} (1,-1/3,6^*) \oplus 
                               \Delta_{d d} (1,2/3,6^*) \, \nonumber \\ & & 
                               \hspace*{6cm}
                               \mbox{under}\quad  \mathcal{G}_{2_L 1_Y 3_C}\, ,
\end{eqnarray}
where the electric charge is expressed in terms of the generators of the SM group and left-right symmetric group as, 
\begin{equation}
Q=T_{3L}+T_{3R}+{\frac{B-L}{2}}=T_{3L}+Y\,.
\label{1}
\end{equation}
Since the fields $\Delta_{\nu \nu} \mbox{(S)}$, $\Delta_{uu}$, $\Delta_{ud}$, $\Delta_{ud}$ and quark fields are mainly 
responsible for non-zero baryon asymmetry and neutron-antineutron oscillation,we need to know the exact 
interactions among them. The desirable interaction Lagrangian for diquark Higgs scalars with the SM quarks at 
TeV scale which will yield observable neutron-antineutron oscillation and post-sphaleron baryogenesis is
\begin{eqnarray}
\hspace*{-0.5cm}\mathcal{L} &\supset& \frac{f_{ij}}{2}\, \Delta_{d d} d_i d_j + \frac{h_{ij}}{2}\, \Delta_{u u} u_i u_j 
             + \frac{g_{ij}}{2\sqrt{2}} \Delta_{ud} \left(u_i d_j + d_i u_j \right) \nonumber \\
            &+& \frac{\lambda}{2} \Delta_{\nu \nu} \Delta_{dd} \Delta_{ud} \Delta_{ud} 
             +  \frac{\lambda^\prime}{2} \Delta_{\nu \nu} \Delta_{uu} \Delta_{dd} \Delta_{dd} + \mbox{h.c.}     \nonumber \\      
&\subset& F\, \left(\psi^T_{R a}\, C^{-1}\, \tau_2\, \vec{\tau} \cdot \Delta^\dagger_{ab}\, \psi_{R b} 
+ \mbox{L}\leftrightarrow \mbox{R} \right) +\mbox{h.c.}\,\, \mbox{under}\quad \mathcal{G}_{224}\, ,
\label{eqn:BVint}
\end{eqnarray}
where $F$, $f,h,g$ are the Majorana couplings and $\tau$ is the generator for $SU(2)$ group.

Within the $SO(10)$ framework, the Yukawa couplings obey the boundary condition, $f_{ij} = h_{ij} = g_{ij}$ in the $SU(2)_L 
\times SU(2)_R \times SU(4)_C \times D$ limit and the same holds true for quartic Higgs couplings $\lambda=\lambda^\prime$ as well. 
All fermions are right-handed (when chiral projection on the operator is suppressed) and a fermion field under the 
high scale Pati-Salam symmetry $\mathcal{G}_{224}$ transforms as, 
\begin{eqnarray}
\psi_{L,R}=\left(
\begin{array}{cccc}
 u_1  & u_2  & u_3 & \nu \\
 d_1  & d_2  & d_3 &  e
\end{array}
\right)_{L,R}
\end{eqnarray} 
The diquark Higgs scalars transforming under the SM gauge group $SU(2)_L \times U(1)_Y \times SU(3)_C$ 
are denoted with quantum numbers as, 
\begin{equation}
\Delta_{\nu \nu} (1, 0, 1),\, \Delta_{u^c u^c} (1, -4/3, 6^*),\, \Delta_{d^c d^c} (1, 2/3, 6^*),\,\, \mbox{and}\,\, 
\Delta_{u^c d^c} (1, -1/3, 6^*)\, . 
\end{equation}
It is clear from eqn\,(\ref{eqn:BVint}) that the Higgs field $\Delta_{\nu \nu} (1, 0, 1) \subset \Delta_R(1, 3, -2, 1) 
\subset (1, 3, \overline{10}) $ is a neutral complex field. The breaking of $\mathcal{G}_{2113} \to \mathcal{G}_{213}$ 
is achieved by assigning a VEV to its neutral component $\Delta_{\nu \nu} \subset \Delta_{R} (1,0,-2,1)$. Its real component 
acquires a VEV in the ground state which can be represented as $\Delta_{\nu \nu} = 
v_{B-L} + \frac{1}{\sqrt{2}} \left(S_r + i \rho \right)$ while the field $\rho$ gets absorbed by the gauge boson corresponding 
to the gauge group $U(1)_{B-L}$. Therefore, the remaining real scalar field $S_r$ is indeed the physical Higgs particle 
which serves our purpose of explaining post-sphaleron baryogenesis and neutron-antineutron oscillation.
\subsection{General expression for CP-asymmetry}
Without loss of generality, if we consider the particle and antiparticle decay modes of $S_r$ ( $S_r$ being its own antiparticle) 
i.e, $S_r \to u^c d^c u^c d^c d^c d^c$ which gives a change of baryon number $\Delta B_{(S_r \rightarrow 6 q^c)} = +2$, and 
$S_r \to \overline{u^c} \overline{d^c} \overline{u^c} 
\overline{d}^c \overline{d}^c \overline{d}^c$ which gives $\Delta B_{(S_r \rightarrow 6 \overline{q^c})} = -2$, 
then the $CP$-asymmetry in baryon number produced by these decays can be quantified as,
\begin{align}
\varepsilon_{CP} &= \frac{\Delta B_{(S_r \rightarrow 6 q^c)}\, \Gamma(S_r \rightarrow 6 q^c)}{\Gamma_\text{tot}}
    +\frac{\Delta B_{(S_r \rightarrow 6 \overline{q^c})}\, \Gamma(S_r \rightarrow 6 \overline{q^c})}{\Gamma_\text{tot}}
    \;,\nonumber\\
 &= \frac{(+2)\, \Gamma(S_r \rightarrow 6 q^c)
 +(-2)\, \Gamma(S_r \rightarrow 6 \overline{q^c})}{\Gamma_\text{tot}} = 2\, \frac{\Gamma -\bar{\Gamma}}{\Gamma_\text{tot}}
    \;,    
\label{eq1:CP_asym_general}
\end{align}
where $\Gamma_\text{tot} = \Gamma +\bar{\Gamma}$ is the total decay rate with $\Gamma \equiv \Gamma(S_r \rightarrow 6 q^c)$ 
and $\bar{\Gamma}\equiv \Gamma(S_r \rightarrow 6 \overline{q^c})$. It is evident from eqn\, (\ref{eq1:CP_asym_general}) 
that we need divergent partial decay rates for particle and antiparticle decays in order to produce correct amount of baryon 
asymmetry and hence we should derive the general conditions under which $\Gamma$ and $\bar{\Gamma}$ can be different. It 
is worth to mention here that the other decay modes of $S_r$ have been ignored for simplicity by adjusting the 
corresponding couplings involved in the respective decay modes.
\begin{figure}[htb!]
\centering
\includegraphics[scale=0.85,angle=0]{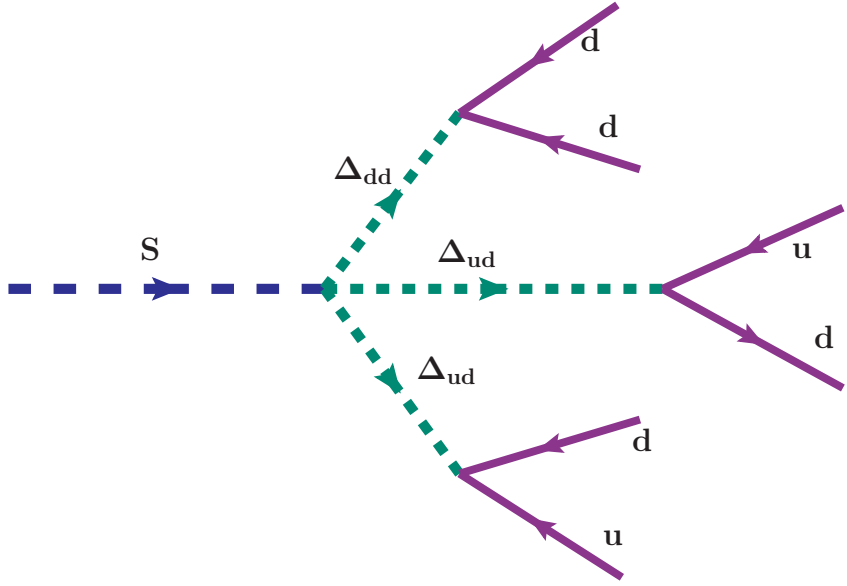}
\caption{Feynman diagram representing the decay of $S\to 6 q$ at tree level in order to explain post-Sphaleron 
         baryogenesis operative at TeV scale. Since $S$ is a real scalar field, the decay mode $S \to 6 \overline{q}$ 
         is possible by reversing the arrow direction of the quark field.}
\label{fig:post-sphaleron:tree}
\end{figure}
In generic situations where the theory is CPT-conserving, there can never be a difference between $\Gamma$ and $\bar{\Gamma}$ 
if one considers only the tree-level process depicted in Fig.~\ref{fig:post-sphaleron:tree} since $\Gamma =\bar{\Gamma}$ at 
tree level. It is found that the nonzero contribution to $\varepsilon_{CP}$ comes from the interference between the tree-level graph 
(shown in Fig.~\ref{fig:post-sphaleron:tree}) and the one-loop corrections (shown in Fig.~\ref{fig:post-sphaleron:loop}).

\subsection{Constraints on post-Sphaleron baryogenesis}
Here we illustrate how post-Sphaleron baryogenesis is slightly different from any other standard baryogenesis 
process. For post-Sphaleron baryogenesis to be successful in explaining the required matter-antimatter 
asymmetry of our Universe, few extra conditions must be satisfied by the model parameters along with the 
\emph{Sakharov conditions} that says, particle interaction must {\bf (i)} violate baryon number, $B$, {\bf (ii)} violate $C$ and $CP$, 
and {\bf (iii)} be out of thermal equilibrium. Firstly, the $S_r$ Higgs scalar should be lighter than other members 
contained in the Pati-Salam multiplet $(1,3, \overline{10})$ i.e, the diquark 
Higgs scalars $\Delta_{qq}$ so that the baryon number conserving decays involving on-shell $\Delta_{qq}$ 
are kinematically forbidden. Secondly, the out of equilibrium baryon number violating decays should 
occur after the electroweak phase transition so that it will not be affected by the Sphaleron processes 
which is proactive at $> $TeV scale. We make it a point here that ref.\,\cite{Babu:2013yca} neatly elaborates 
the mechanism of post-sphaleron baryogenesis.
\subsection{Out of equilibrium condition}
For effectively creating the baryon asymmetry of the universe via post-Sphaleron baryogenesis, the decays 
of $\Gamma(S_r \rightarrow 6 q^c)$  should satisfy the out of equilibrium condition, which is described 
by $\Gamma_{S_r} \lesssim H(T)$ where $\Gamma= \Gamma(S_r \rightarrow 6 q^c)= \frac{36}{\left(2 \pi \right)^9} 
\frac{\left(\textbf{Tr}[f^\dagger f] \right)^3\, \lambda^2 M^{13}_S}{6 M^{12}_\Delta}$ is the total decay width 
and $ H \simeq 1.66 \sqrt{g^*_s}\, \frac{T^2}{M_\text{Pl}} \;,$ is the Hubble parameter with the reduced Planck 
mass $M_{\textrm{Pl}} \simeq 1.2 \times 10^{18}\,\textrm{GeV}$ and $g^*_s$ is the number of relativistic degrees 
of freedom. In order to satisfy the out of equilibrium condition, we should have 
\begin{eqnarray}
& &\Gamma_{S_r} \simeq H \left|_{(T=T_d)}\right. \nonumber\\
&\Rightarrow& \mbox{T}_d =\bigg[ \frac{36\,\lambda^2\,\left(\textbf{Tr}[f^\dagger f] \right)^3\,M_{\textrm{Pl}} M^{13}_S}{ 
\left(2 \pi\right)^9\, 1.66\, g^{1/2}_{\ast} \left(6 M_\Delta \right)^{12} } \bigg]^{1/2}  \simeq 6.1 \times 
\left(\frac{M^{13}_S}{M^{12}_{\Delta}}\right)^{1/2} \,\mbox{GeV}^{1/2}
\label{eq2:Sdecay-tree}
\end{eqnarray}
To illustrate the mechanism of post-sphaleron baryogenesis, we require extra fields $\Delta_{uu}$, $\Delta_{ud}$ and 
$\Delta_{dd}$ as color sextets and $SU(2)_L$ singlet scalar bosons that couple to the right-handed quarks contained 
in the Pati-Salam multiplet $(1,3,\overline{10})$. For set of model parameters $M_S=500$ GeV, $M_\Delta \simeq 1000$ GeV, 
the decoupling temperature is found to be 2 GeV which is well below the EW scale where the Sphaleron has been decoupled. 
Hence, it is inferred from the above equation that the decay of $S$ goes out of equilibrium around $T\simeq M_S$. Below this 
temperature ($T < M_S$), the decay rate falls very rapidly as the temperature cools down.

\begin{figure}[h!]
\centering
\includegraphics[scale=0.95,angle=0]{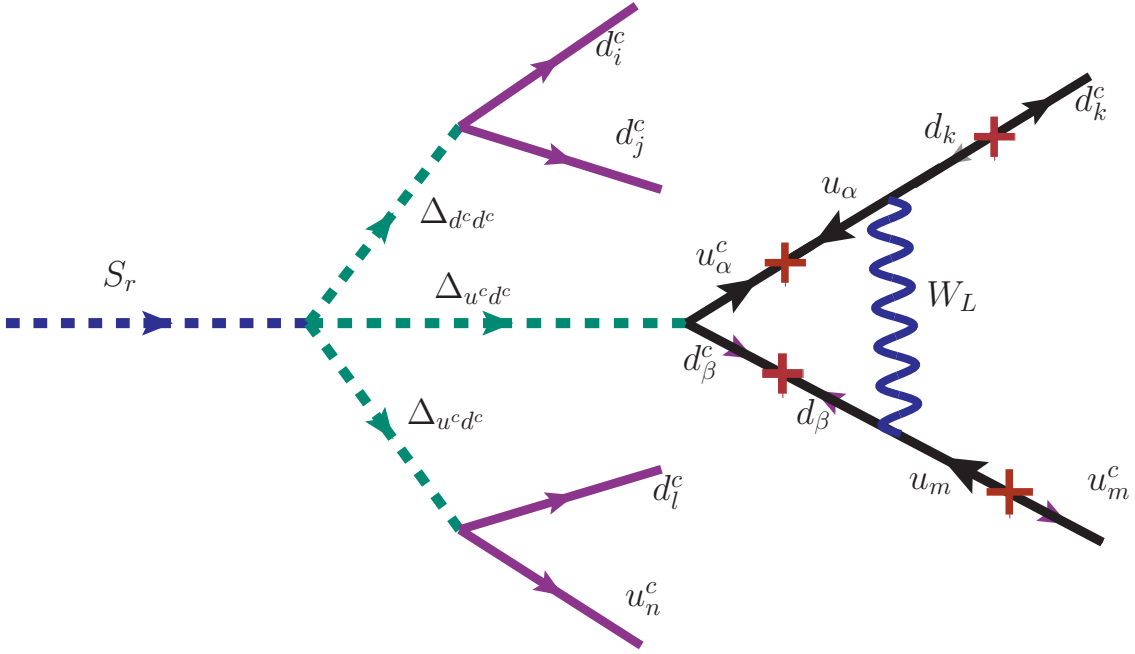}
\caption{Feynman graphs of the one-loop vertex correction for $\Gamma(S_r \rightarrow 6 q^c)$.}
\label{fig:post-sphaleron:loop}
\end{figure}
\subsection{Estimation of net baryon asymmetry}
Now we concentrate on estimating the CP-asymmetry coming from the interference term between the tree level and the one-loop level 
diagrams for the decay of $S_r$ which is shown in Fig.\ref{fig:post-sphaleron:tree} and Fig.\ref{fig:post-sphaleron:loop} 
respectively. For discussion on baryon number violation in the loop diagram and necessary derivation of the interference diagram, 
interested readers may go through reference \cite{Babu:2013yca}. In the present work, we only check whether or not the 
representative set of model parameters provide the correct number for the required baryon asymmetry of the universe. 
Hence, without going deep into the derivation, we simply note here down, the calculated CP-asymmetry 
for post-sphaleron baryogenesis via decay of $S_r$ with baryon number violating interactions. 
\begin{align}
\varepsilon_{\rm wave} &\simeq \frac{g^2}{64 \pi \mbox{Tr}(f^\dagger f)} f_{j \alpha} V^*_{j \beta} f_{i \alpha} \delta_{i3}
    \frac{m_t m_j}{m^2_t-m^2_j} \sqrt{\left(1-\frac{m^2_W}{m^2_t} +\frac{m^2_\beta}{m^2_t}\right)^2- 4 \frac{m^2_\beta}{m^2_t}} 
    \nonumber \\
   & \times \bigg[2 \left(1-\frac{m^2_W}{m^2_t} +\frac{m^2_\beta}{m^2_t}\right) 
    + \left(1+\frac{m^2_\beta}{m^2_t}\right)\left(\frac{m^2_t}{m^2_W} +\frac{m^2_\beta}{m^2_t}-1\right) 
    - 4 \frac{m^2_\beta}{m^2_W}\bigg]    \;,  
    \label{eq3:wave-cpasy}\\
\varepsilon_{\rm vertex} &\simeq \frac{g^2}{32 \pi \mbox{Tr}(f^\dagger f)} f_{i \beta} V^*_{i \beta} f_{i \alpha} \delta_{i3} 
    \frac{m_j m_\beta}{m^2_W} \bigg[1+\frac{9 m^2_W}{M^2_S}\mbox{ln}\left(1+\frac{M^2_S}{3 m^2_W} \right) \bigg] \;,
    \label{eq3:vertex-cpasy} \\
\varepsilon_{\rm CP} &=\varepsilon_{\rm wave} + \varepsilon_{\rm vertex} \, .
\label{eq:CP_asym_expressn}
\end{align}
Here the expression in eq.(\ref{eq3:wave-cpasy}) represents the CP-asymmetry coming from interference between the tree and 
one-loop self energy diagram while the expression in eq.(\ref{eq3:vertex-cpasy}) represents the CP-asymmetry due to interference 
of the tree and one-loop vertex diagram (see ref.\cite{Babu:2013yca} for details). 
In the above expression, $V$ is the well known CKM matrix in the quark sector, $i,j$ correspond to the up-quark indices 
$u,c,t$ while $\alpha,\beta$ represent to down-quark indices $d,s,b$. Sum over repeated indices (Einstein convention) 
is implicitly assumed here. The $\delta_{i3}$ is due to the fact that the CP asymmetry is non-zero only when we have 
a top quark in the final state (since only the CKM elements involving third generation have a large imaginary part). 

\begin{figure}[t]
\centering
\includegraphics[scale=1.2,angle=0]{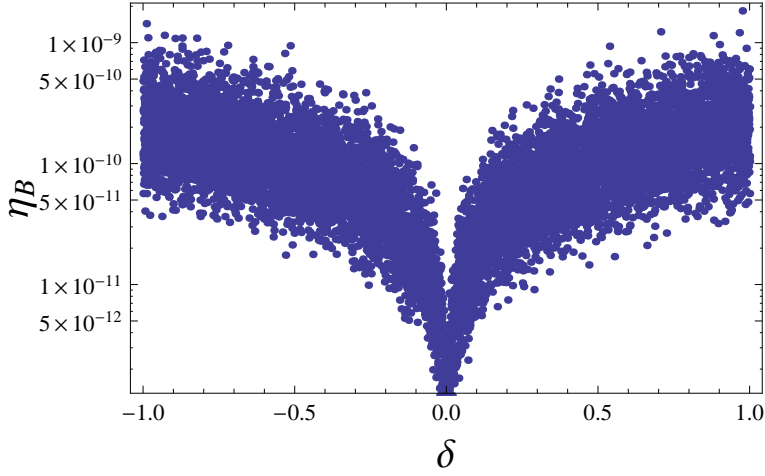}
\caption{Estimation of final baryon asymmetry in terms of CP-asymmetry with overall phase $\delta$ contained 
         in the CKM mixing matrix.}
\label{plot:CP-asymmetry}
\end{figure}

As mentioned earlier,the mechanism of post-sphaleron baryogenesis provides a natural explanation for the observed baryon 
asymmetry of our universe i.e, $\eta_B\simeq 10^{-10}$. Using $m_c=1.27$ GeV, $m_b=4.25$ GeV, $m_t=172$ GeV, CKM mixing 
elements $V_{CKM}$ and Yukawa couplings relevant for color scalar particles in their allowed range, the CP-asymmetry 
via the decay of $S_r$ through loop diagrams with the exchanges of $W^{\pm}$ bosons is estimated to be $10^{-8}$. A  
further dilution of the baryon asymmetry arises from the fact that $T_d \ll M_S$, since the decay of $S_r$ releases entropy 
into the universe. As a result the final baryon asymmetry, taking into account the dilution factor, becomes
\begin{eqnarray}
\eta_B = \varepsilon_{\rm CP} \times \left(\frac{T_d}{M_S} \right)\, , 
\end{eqnarray}
where $T_d$ is the decoupling temperature of the color scalar and $M_S$ is the mass of the scalar. The condition $T_d/M_S \geq 10^{-2}$, 
otherwise leads to suppressed baryon asymmetry, which finally results a baryon asymmetry in the range of $10^{-10}$. 
The scatter plot between the final baryon asymmetry including dilution factor ($\eta_B$) 
with this phase ($\delta_{i3}$) is shown in Fig.\,\ref{plot:CP-asymmetry}. 

\section{OBSERVABLE NEUTRON-ANTINEUTRON OSCILLATION WITH TEV SCALE DIQUARK HIGGS SCALARS:}
\begin{figure}[h!]
\centering
\includegraphics[scale=0.55,angle=0]{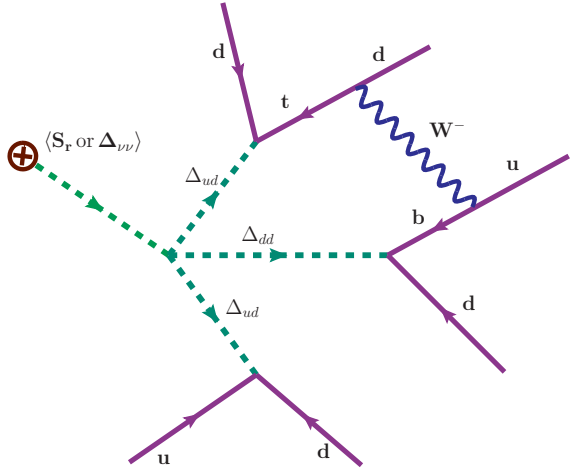}
\caption{Loop contributions to neutron-antineutron oscillation in the post-Sphaleron baryogenesis operative 
         at TeV scale.}
\label{fig:nnbar-loop}
\end{figure}
\subsection{Feynman amplitudes for neutron-antineutron oscillation}
We consider the contributions arising only from the RH diquark Higgs fields having masses at TeV scale while ignoring the 
contributions from LH diquark Higgs fields since they have masses at around $\mbox{eV}$ 
range. The Feynman diagrams contributing to the neutron-antineutron oscillation are shown in Fig.\,\ref{fig:nnbar-loop} (loop-diagram), 
Fig.\,6({\bf a}) and Fig.\,6({\bf b}). Our prime goal is to estimate the mixing time 
for this loop diagram, clarifying why we have suppressed other contributions within our model parameters.

There are two types of contributions to $n-\overline{n}$ oscillation in the right-handed 
sector at loop level (i) one involving one $u^c u^c$-type and two $d^c d^c$-type, (ii) 
other one involving one $d^c d^c$-type and two $u^c d^c$-type $\Delta$-bosons. The Feynman 
amplitude for the second type of contribution where one needs to change the two $b^c$ quarks 
to two $d^c$ quarks from the already generated effective operator  $u^c d^c b^c u^c d^c b^c$ 
via a second order weak interactions (given in Fig.\,\ref{fig:nnbar-loop}) can be written as,
\begin{eqnarray}
\mathcal{A}^{\rm 1-loop}_{n-\overline{n}} \simeq 
       \frac{\left(f_{ud}\right)_{11} \left(f_{ud}\right)_{13} \left(f_{dd}\right)_{13}\, \lambda v_{B-L}}
       {M^4_{u^c d^c} M^2_{d^c d^c}}\, \frac{g^4\, V^2_{td}\, m^2_b\,m^2_t}{\left(16 \pi^2 \right)^2 M^4_{W_L}}\,
       \mbox{\Large log}\left(\frac{m^2_b}{M^2_{W_L}}\right)
\label{eq:ampl-loop}
\end{eqnarray}
And, the Feynman amplitude for tree level processes shown in Fig.\,6({\bf a}) and Fig.\,6({\bf b}) 
(which are suppressed with the choice of our model parameters), can be written as,
\begin{eqnarray}
\mathcal{A}^{\rm tree}_{n-\overline{n}} &=& \mathcal{A}^{\rm (a)}_{n-\overline{n}}+ \mathcal{A}^{\rm (b)}_{n-\overline{n}} \nonumber \\
       &\simeq&\frac{\left(f_{dd}\right)_{11} \left(f_{ud}\right)^2_{11}\, \lambda\, v_{B-L}}
       {M^4_{u^c d^c} M^2_{d^c d^c}} + \frac{\left(f_{uu}\right)_{11} \left(f_{dd}\right)^2_{11}\, \lambda\, v_{B-L}}
       {M^4_{d^c d^c} M^2_{u^c u^c}}
\label{eq:ampl-tree}
\end{eqnarray}

\begin{figure*}[htb]
\begin{minipage}[t]{0.49\textwidth}
\hspace{-0.4cm}
\begin{center}
\includegraphics[scale=0.43,angle=0]{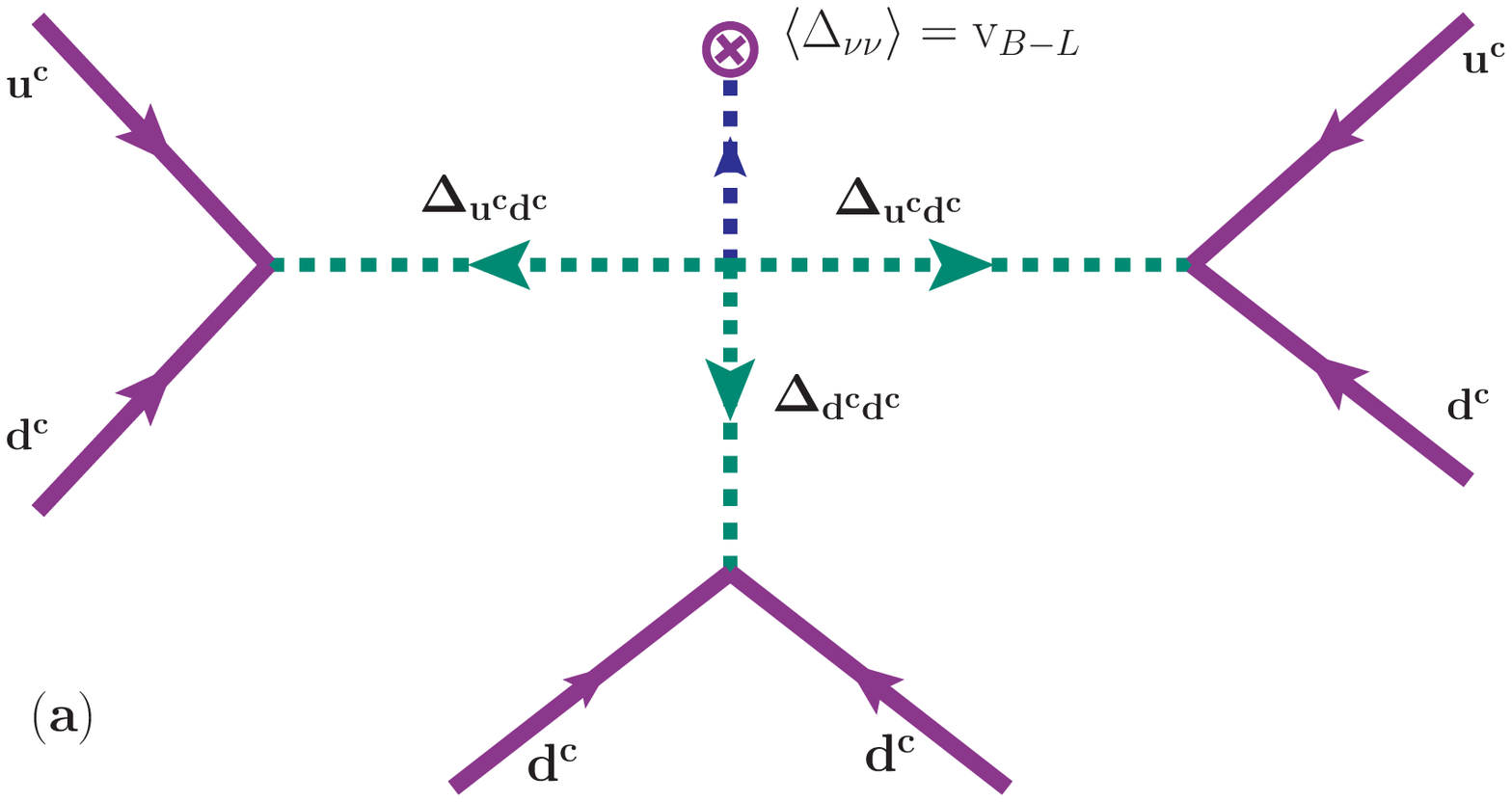}
\end{center}
 \end{minipage}
 \hfill
 \begin{minipage}[t]{0.49\textwidth}
 \hspace{-0.4cm}
 \begin{center}
 \includegraphics[scale=0.43,angle=0]{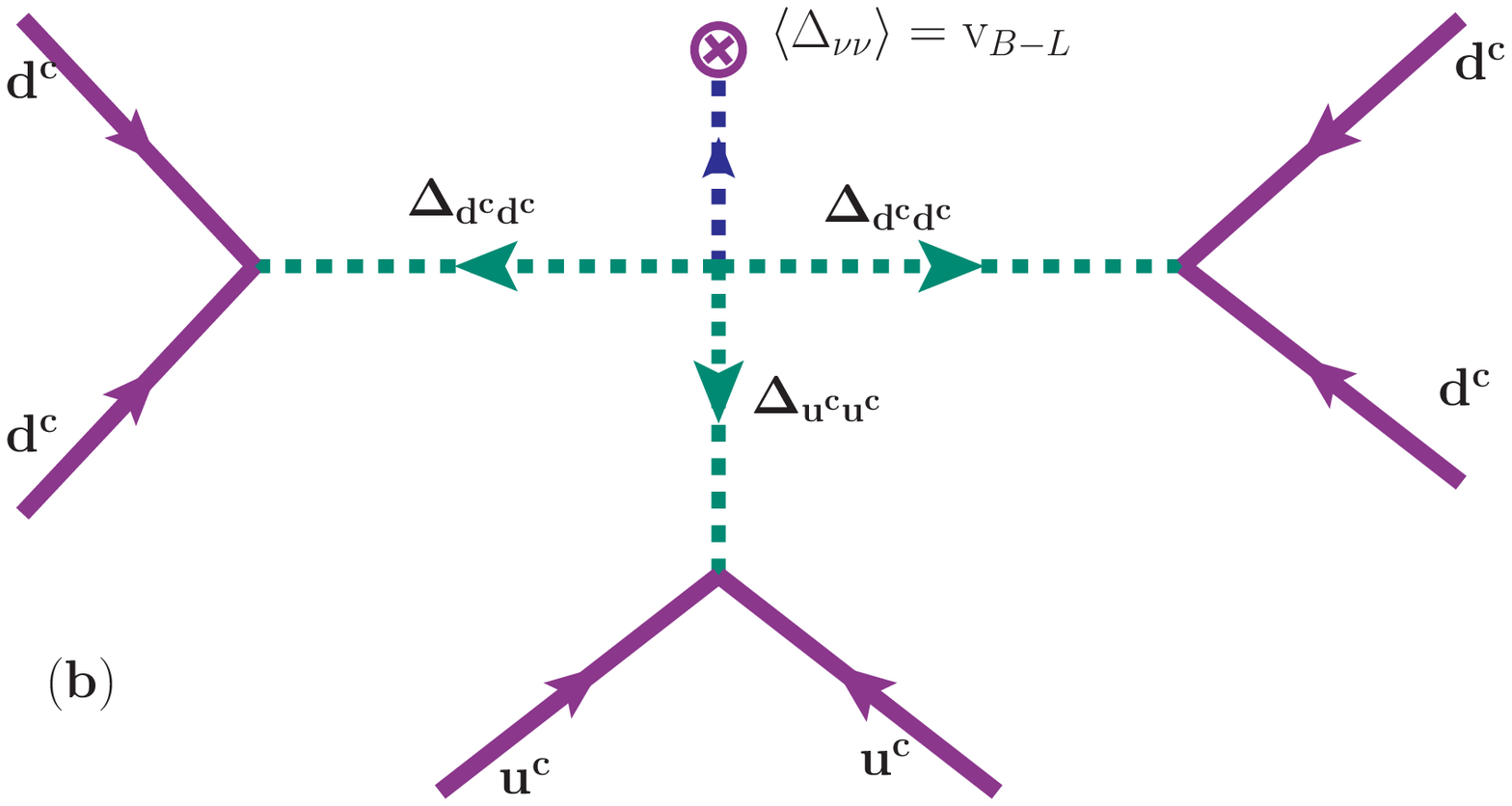}
 \end{center}
 \end{minipage}
 \caption{Feynman diagrams contributing to neutron-antineutron oscillation. The figure in {\it left-panel} 
 involves two $\Delta_{u^c d^c}$ and one $\Delta_{d^c d^c}$ bosons whereas the figure in {\it right-panel} involves 
 two $\Delta_{d^c d^c}$ and one $\Delta_{u^c u^c}$ bosons. The structure of the theory is such that these tree-level 
 contributions are suppressed in the present work.}
  \label{fig:supp-nnbar}
\end{figure*}

\subsection{Prediction for neutron-antineutron mixing time $\tau_{n-\overline{n}}$}
Before estimating the  $n-\overline{n}$ oscillation mixing time one should carefully fix the input parameters 
in order to satisfy flavor changing neutral current (FCNC) constraints and to give correct amount of baryon 
asymmetry of the universe. For example, using diquark sextet Higgs scalar mass around TeV scale, the corresponding 
Yukawa coupling $(f_{dd})_{11}\simeq 0.001-0.1$ along with other allowed range of model parameters contradicts  
the FCNC constraints and hampers post-sphaleron baryogenesis even though it predicts neutron-antineutron oscillation 
time (as shown in Fig.\,6) within the experimental search limits. So this means that one has to choose the Majorana Yukawa coupling 
$f$ accordingly. Now we briefly discuss how this choice of $f$ can be achieved within the framework of SO(10) (elaborated in 
ref \cite{Awasthi:2013ff}). 

It is found in ref \cite{Awasthi:2013ff} that all charged fermion masses and CKM mixing can 
be fitted well at GUT scale within the framework of SO(10) with two kinds of structures; I) with single Higgs representation 
$126_H$, II) with two Higgs representations $126_H$, $126_H^\prime$. As it has been derived, structure-I with Yukawa coupling 
$f_{126_H}$ = diag(0.0236, -0.38, 1.5) estimates $n-\bar{n}$ oscillation mixing time to be $10^9$ secs which doesn't 
serve our purpose. Rather we consider structure-II where the dominant 
contribution to $n-\overline{n}$ oscillation comes from the loop diagram while suppressing the tree level contribution. 
This choice of having two Higgs $126_H$, $126_H^\prime$ not only fits fermion masses at GUT scale, but also allows RH neutrino 
Majorana mass and hence corresponding Yukawa coupling $f_{126_H^\prime}$ as per our requirement.
Due to the second Higgs representation ${126}^{\prime}$ with its Yukawa coupling $f^{\prime}$ to fermions 
we get $v_{\xi^{\prime}}= 1-100 $ MeV following the same procedure, provided all other components are at the 
GUT scale except $\xi^{\prime}(2,2,15)$ which is at the intermediate scale $M_{\xi^{\prime}}=10^{13}-10^{14}$ GeV. 
By treating the mass of $\xi(2,2,15) \subset 126$ to remain at its natural GUT-scale  value, its induced VEV is 
negligible and precision unification with large GUT scale value is unaffected except for phenomenologically 
inconsequential additional threshold effects. Then defining $F= f^{\prime}v_{\xi^{\prime}}$ gives exactly 
the same fit to the GUT scale fermion masses and mixings but now  with the diagonal structure $f_i^{\prime} 
=(0.0236, -0.38, 1.5)$. But since $<\Delta_R^{\prime}>=0$ and only $\Delta_R \subset {126}_H$ with VEV $v_R$ 
is used to break $G_{2113}$, the coupling $f$ and hence $M_N$ are allowed to have any $3 \times 3$ form 
without any restriction. In order to suppress the tree level contributions to $n-\overline{n}$ oscillation 
as shown in Fig.\,6 which otherwise causes problem in baryon asymmetry, we particularly choose 
the Majorana coupling $f_{dd}$ as per our requirement, i.e, ${f_{dd}}_{11}\leq 10^{-5}$.

\begin{table}[h!]
\begin{center}
\begin{tabular}{|c|c|c|c|c|c|c|}
\hline
$f_{13}$ & $g_{11}$ & $g_{13}$ & $\lambda$ & ${M_\Delta}_{ud}$ (GeV) & ${M_\Delta}_{dd}$ (GeV) & $\tau_{n-\bar{n}}$ (sec)\\
\hline
0.001 & 0.01 & 0.01 & 0.1 & $10^3$ & $10^4$ & $3.96\times 10^8$\\
\hline
0.001 & 0.01 & 0.01 & 0.1 & $10^3$ & $10^5$ & $8.72\times 10^{10}$\\
\hline
0.001 & 0.01 & 0.01 & 1 & $10^3$ & $10^5$ & $3.29\times 10^9$\\
\hline
0.001 & 0.001 & 0.001 & 0.1 & $10^3$ & $10^4$ & $4.42\times 10^{10}$\\
\hline
\end{tabular}
\caption{Numerical estimation of neutrino-antineutrino oscillation time}
\label{tab:nnbar-mixing}
\end{center}
\end{table}

\begin{figure}[htb!]
\centering
\includegraphics[scale=1.1,angle=0]{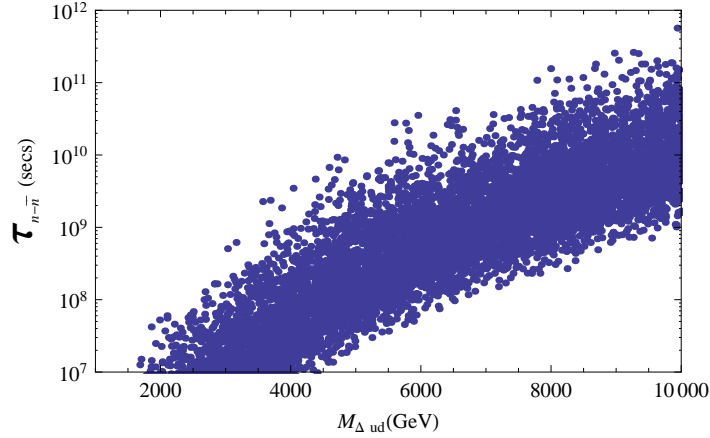}
\caption{Estimation of $\tau_{n-\bar{n}}$ as a function of di-quarks mass $M_{\Delta_{ud}}$.}
\label{plot:nnbar-Mud}
\end{figure}

\begin{figure}[htb!]
\centering
\includegraphics[scale=1.1,angle=0]{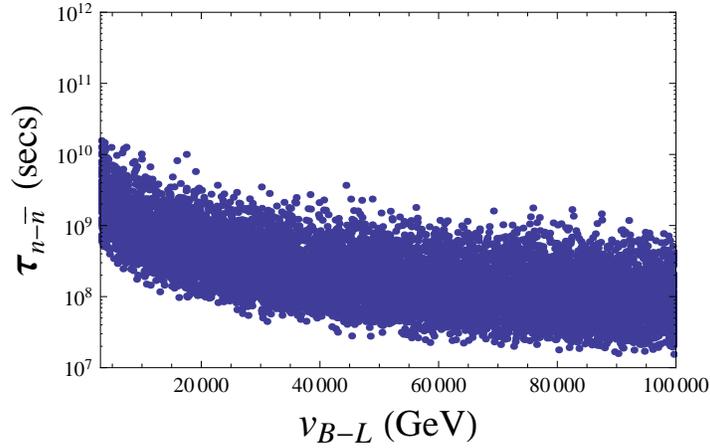}
\caption{Estimation of $\tau_{n-\bar{n}}$ as a function of $B-L$ breaking scale $v_{B-L}$ 
while keeping other model parameters within their allowed range consistent with mechanism 
of post-sphaleron baryogenesis.}
\label{plot:nnbar-vBL}
\end{figure}

Using this particular choice of Yukawa couplings i.e, ${f_{dd}}_{11},\, {f_{dd}}_{22},\leq 10^{-5}$ and others in 
the range of $0.001-1.0$, one can calculate the mixing time for neutron-antineutron oscillation as a function of 
Mass of color Higgs scalar ($B-L$ breaking scale) as shown in Fig.\,\ref{plot:nnbar-Mud} (Fig.\,\ref{plot:nnbar-vBL}).

The $n-\bar{n}$ amplitude can be translated into the $n-\overline{n}$ oscillation time as, 
\begin{equation}
 \tau_{n-\bar{n}}^{-1} = {\delta m}_{n-\bar{n}} = C_{{\rm \small QCD}} (\mu_\Delta, \mbox{1\,GeV}) |A_{n-\bar{n}}^{\rm 1-loop}|  
\end{equation}
with $C_{{\rm \small QCD}} (\mu_\Delta, \mbox{1\,GeV})$ = $0.1 \mbox{GeV}^6$ as used in ref.\cite{Babu:2013yca}. The estimated 
$n-\bar{n}$ oscillation time for various choice of model parameters i.e, ${f_{ud}}_{11} \leq {10}^{-5}$, $M_S = (100-5000)GeV$, 
B-L breaking scale from (3-5)TeV and the masses of ${M_\Delta}_{ud/dd}$ between $M_S$ and $V_{B-L}$, $\lambda\simeq 0.01-1.0$ 
is presented in Table.\ref{tab:nnbar-mixing}.

\subsection{Coupling Unification including diquarks at TeV scale}
It is prominent that the post-sphaleron baryogenesis and neutron-antineutron oscillation phenomena require existence of 
color Higgs scalars, having masses around TeV scale. In this subsection, we intend to examine whether unification of gauge couplings 
is still possible after the addition of extra color scalars $\Delta_{ud}$, $\Delta_{dd}$, $\Delta_{uu}$ to the existing particle 
content as noted in Sec.\ref{sec:rge}, by studying their respective renormalization group equations.  
The one-loop beta coefficients derived for the present model along with their gauge 
symmetry groups, range of mass scales and spectrum of Higgs scalars necessary for gauge 
coupling unification to explain TeV scale post-sphaleron baryogenesis and 
neutron-anti-neutron oscillation are given below
\begin{eqnarray}
& &{\bf \mbox{(i)}\, \mu=M_Z (91.817\, \mbox{GeV}) - M_T (1\, \mbox{TeV}) }: \quad \mathcal{G} = \mathcal{G}_{2_L 1_Y 3_C}\equiv \mbox{SM}, 
       \nonumber \\ 
       & &\mbox{Higgs:}\, \Phi(2, 1/2, 1)_{10}:\quad \quad 
       \pmb { a_i}=\left(-19/6,\, 41/10,\, -7\right)\,; \\
& &{\bf \mbox{(ii)}\, \, \mu=M_T(1\, \mbox{TeV}) - M_{B-L}(3\, \mbox{TeV}):} \quad \mathcal{G} = \mathcal{G}_{2_L 1_Y 3_C} 
    \nonumber \\
    & &\mbox{Higgs:}\, \Phi(2,1/2,1)_{10} \oplus  S(1, 0, 1)_{126}\subset \Delta_R \oplus \Delta_{u^c d^c} (1, -1/3, 6^*)_{126} 
                     \oplus \Delta_{d^c d^c} (1, 2/3, 6^*)_{126}\, \nonumber \\
                   & &\hspace*{1.2cm} \oplus \Delta_{u^c u^c} (1, -4/3, 6^*)_{126}: \nonumber \\
    & &\pmb{ a_i} = \left(-19/6,\, 207/30,\, -27/6\right) 
\end{eqnarray}
\begin{eqnarray}
& &{\bf \mbox{(iii)}\, \, \mu=M_{B-L}(3\, \mbox{TeV}) - M_\Omega(10\, \mbox{TeV}):} 
                          \quad \mathcal{G} = \mathcal{G}_{2_L 1_R 1_{B-L} 3_C} \nonumber \\
      & &\mbox{Higgs:}\,\Phi_1(2, 1/2, 0, 1)_{10} \oplus \Phi_2(2, -1/2, 0, 1)_{10^\prime} \oplus \Delta_R(1, 1, -1, 1)_{126} 
                       \oplus \chi_R(1, 1/2, -1/2, 1)_{16},\,  \nonumber \\ 
      & &\hspace*{1.2cm} \oplus \Delta_{u^c d^c}(1, 1, -2/3, 6^*)_{126} \oplus \Delta_{d^c d^c} (1, 0, -2/3, 6^*)_{126} 
                   \oplus \Delta_{u^c u^c} (1, 0, -2/3, 6^*)_{126}: \nonumber \\
& &\pmb{a_i} = \left(-3,\, 35/4,\, 45/8, -27/6\right) 
\end{eqnarray}
\begin{eqnarray}
& &\hspace*{-0.4cm} {\bf \mbox{(iv)}\, \, \mu=M_\Omega(10^{4}\, \mbox{GeV}) - M_C(10^{5}-10^{6}\, \mbox{GeV}):} 
           \quad \mathcal{G} = \mathcal{G}_{2_L 2_R 1_{B-L} 3_C} \nonumber \\
           & &\mbox{Higgs:}\,\Phi_1(2, 2, 0, 1)_{10} \oplus \Phi_2(2, 2, 0, 1)_{10^\prime} \oplus \Delta_R(1, 3, -1, 1)_{126}  
                      \oplus  \chi_R(1, 2, -1/2, 1)_{16},\, \nonumber \\
           & &\hspace*{1.2cm} \oplus \Delta_{u^c d^c} (1, 3, -2/3, 6^*)_{126} \oplus \Delta_{d^c d^c} (1, 3, -2/3, 6^*)_{126} 
                      \oplus \Delta_{u^c u^c} (1, 3, -2/3, 6^*)_{126} \nonumber \\
           & &\hspace*{1.2cm} \oplus \Omega_R(1, 3, 0, 1)_{210} \nonumber \\
& &\pmb { a_i} = \left(-8/3,\, 4/3,\, 55/4, -2\right) 
\end{eqnarray}
In analogy to the above discussion, we have two scenarios; one without bitriplet and another with bitriplet Higgs scalar 
(3,3,1) under the Pati-Salam group $SU(2)_L \times SU(2)_R \times SU(4)_C$ while its effect has been included from $M_C$ 
onwards to the unification scale $M_U$. Accordingly, we have estimated the one-loop beta coefficients for these two 
scenarios as 
\begin{eqnarray}
& &\hspace*{-0.4cm} {\bf \mbox{(v)}\, \, \mu=M_C - M_\xi:} \quad \mathcal{G} = \mathcal{G}_{2_L 2_R 4_C} \nonumber \\
     & &\mbox{Higgs:}\,\Phi_1(2, 2, 1)_{10} \oplus \Phi_2(2, 2, 1)_{10^\prime} \oplus \Delta_R(1, 3, \overline{10})_{126} 
                       \oplus \chi_R(1, 2, \overline{4})_{16} \oplus \Omega_R(1, 3, 15)_{210} \nonumber \\
& &\pmb{a_i} = \left(-8/3,\, 29/3,\, -14/3\right)
\end{eqnarray}
\begin{eqnarray}
& &\hspace*{-0.4cm} {\bf \mbox{(vi)}\, \, \mu=M_\xi - M_P:} \quad \mathcal{G} = \mathcal{G}_{2_L 2_R 4_C} \nonumber \\
& &\mbox{Higgs:}\,\Phi_1(2, 2, 1)_{10},\, \Phi_2(2, 2, 1)_{10^\prime},\,\Delta_R(1, 3, \overline{10})_{126},\,  
               \chi_R(1, 2, \overline{4})_{16}, \Omega_R(1, 3, 15)_{210} + \xi(2, 2, 15)_{126^\prime} \nonumber \\
& &\pmb { a_i} = \left(7/3,\, 44/3,\, 2/3\right) 
\end{eqnarray}
\begin{eqnarray}
& &\hspace*{-0.4cm} {\bf \mbox{(vii)}\, \, \mu=M_P - M_U:} \quad \mathcal{G} = \mathcal{G}_{2_L 2_R 4_C} \nonumber \\
& &\mbox{Higgs:}\,\Phi_1(2, 2, 1)_{10},\, \Phi_2(2, 2, 1)_{10^\prime},\, \Delta_R(1, 3, \overline{10})_{126},\,  
           \Delta_L(3, 1, 10)_{126},\, \chi_R(1, 2, \overline{4})_{16},\, \chi_L(1, 2, 4)_{16},\, \nonumber \\
         & &\quad \quad \quad \Omega_R(1, 3, 15)_{210},\, \Omega_L(3, 1, 15)_{210},\, 
\xi(2, 2, 15)_{126^\prime},\, \Sigma^\prime(1,1,15)_{210},\,  \nonumber \\
& &\pmb { a_i} = \left(44/3,\, 44/3,\, 6 \right) \nonumber \\
\end{eqnarray}
\begin{figure}[htb!]
\centering
\includegraphics[scale=0.98,angle=0]{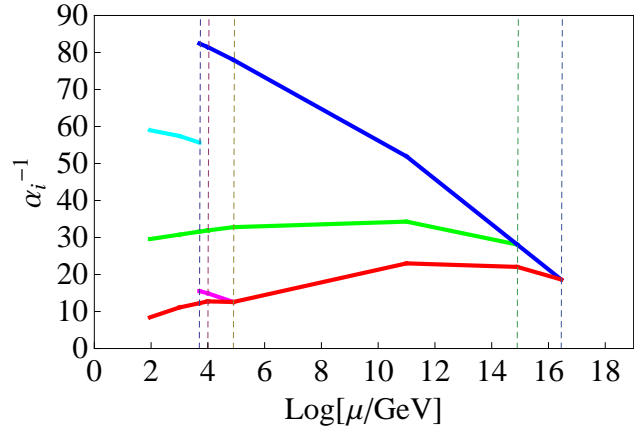}
\caption{Coupling unification for the present model where $\Delta_{u^c d^c}$, $\Delta_{d^c d^c}$, and $\Delta_{u^c u^c}$ 
         have been included at TeV scale keeping in mind that these particle mediate neutron-antineutron oscillation 
         and baryon asymmetry and including $\xi(2,2,15)$ around $10^{12}$ GeV in order to fit the fermions 
         masses at GUT scale.}
\label{fig:unif}
\end{figure}
The gauge coupling unification after the addition of extra color sextet scalars particles is shown in Fig.\,\ref{fig:unif} 
with the allowed mass scales desirable for our model predictions,
\bea 
& &M_{B-L}= 4-7~{\rm TeV},\, M_\Omega= 10~{\rm TeV},\, M_C = 10^{5}-10^{6}\, \mbox{GeV}\, ,\nn \\
& &M_P\simeq 10^{14.65}~{\rm GeV\,\, and\,\,} M_{\rm U}\simeq 10^{16.25}~{\rm GeV} \, .
\eea
\section{VIABILITY OF THE MODEL}
As already known, the lepton flavor and lepton number violating dilepton signals can be probed 
from the production of heavy RH Majorana neutrino via $p+p \to W^\pm_R \to \ell^\pm_\alpha +N_R$, 
from which $N_R$ can be further decayed into $N_R \to W^*_R \to \ell^{\mp}_\beta = 2j$. This process, 
being the main channel for $N_R$ production via on-shell $Z_R$ production and $W_R$ fusion, needs to be 
verified at LHC and our model suits the purpose, since we have $W_R$, $Z_R$ gauge bosons and scalar 
diquarks at TeV scale. A more pleasant situation is that the model, though non-supersymmetric, predicts 
similar branching ratios as in supersymmetric models for LFV processes like $\mu \to e\gamma$, $\tau \to 
\mu\gamma$, and $\tau \to e\gamma$. And the predicted branching ratios for these LFV decays, being closer 
to the current experimental search limits can be used to verify the left-right framework in this model. 
Moreover the estimated neutron-antineutron oscillation mixing time, gauge coupling unification and  
proton life time in the model stay in the range of ongoing search experiments.

Besides all these points, the model can also predict a number of verifiable new physical quantities like 
(i) new non-standard contribution to $0\nu 2\beta$ rate in the $W_L-W_L$ channel, (ii) contributions to branching 
ratios of lepton flavor violating (LFV) decays, (iii) leptonic CP-violation due to non-unitarity effects, (iv) 
experimentally verifiable proton decay modes such as $p\to e^+\pi^0$, provided the gauged inverse seesaw mechanism 
is found to be operative. We find it appropriate to mention here that these physical quantities were also discussed 
in a recent work \cite{Awasthi:2013ff}, but in that model the asymmetric left-right gauge symmetry was incorporated 
at $\simeq 10$ TeV.

\section{CONCLUSION}
We have closely studied the mechanism of post-sphaleron baryogenesis, that can potentially explain matter-antimatter 
asymmetry of the present universe, by analyzing the basic interactions using quarks and diquark Higgs scalars under 
high scale Pati-Salam symmetry and low scale SM like interactions at TeV scale. The study estimates the total baryon 
asymmetry to be $\eta_B \simeq {\cal O}(10^{-10})$ and neutron-antineutron oscillation with mixing time to be 
$\tau_{n-\bar{n}}\simeq {\cal O}(10^{-10} -10^{-8})$ secs which can be accessible at ongoing search experiments. 
We have made an humble attempt to embed the framework of PSB in a non-SUSY $SO(10)$ model with Pati-Salam symmetry as a  
low scale intermediate breaking step 
where we have shown a strong interlink between post-sphaleron baryogenesis 
and neutron-antineutron oscillation operative at TeV scale and laid out a novel mechanism of inducing required CP-asymmetry via the SM 
$W_L^{\pm}$ loops.

More essentially, we have embedded TeV scale LR model 
within the framework of  $SO(10)$ model where the predicted mass for light neutrinos matches with the neutrino oscillation 
data. 
Our calculations indicate that TeV scale masses of 
$W^\pm_R$ and heavy RH neutrinos can also give dominant non-standard contributions to neutrinoless double beta decay which 
may sound crucial to the experimentalists. Some more good features of the model are explanation of non-zero light neutrino masses 
via extended/inverse seesaw mechanism, new non-standard contribution to neutrinoless double beta decay, leptonic CP-violation from 
non-unitary effects.

\noindent
{\bf ACKNOWLEDGEMENT}

\noindent Sudhanwa Patra would like to thank the organizers of workshop entitled \textquotedblleft 
Majorana to LHC: Origin of neutrino Mass\textquotedblright  at ICTP, Trieste, Italy during 2-5 October, 2013 
where the idea for this work was conceived. Both the authors sincerely acknowledge P.S. Bhupal Dev for his 
useful clarification while preparing the manuscript. Prativa Pritimita is grateful to the Department 
of Science and Technology, Govt. of India for INSPIRE Fellowship (IF140299). The work of Sudhanwa Patra 
is supported by the Department of Science and Technology, Govt. of India under the financial grant 
SERB/F/482/2014-15.


\begin{thebibliography}{50}
\bibitem{Dunkley:2008}
J. Dunkley et al. ``{\em { {\bf WMAP Collaboration}}},''. 
  \href{http://arxiv.org/abs/0803.0586}{arXiv:0803.0586 [astro-ph]}.
\bibitem{Komatsu:2011}  
Komatsu, E. et al ``{\em { Seven-Year Wilkinson Microwave Anisotropy 
   Probe (WMAP) Observations: Cosmological Interpretation}},''.
  \href{http://dx.doi.org/10.1088/0067-0049/192/2/18}{Astrophys. J. Suppl. {\bf 192} (2011) 18}
  \href{http://arxiv.org/abs/1001.4538}{arXiv:1001.4538 [astro-ph]}.
\bibitem{Yao:2006}
W.-M. Yao et al. ``{\em {{\bf Particle Data Group:2006}, partial update for edition 2008 
   (URL: http://pdg.lbl.gov)}},''
   \href{http://dx.doi.org/J. Phys. G 33, 1 (2006)}{J. Phys. {\bf G\,33}, 1 (2006)}.
\bibitem{Fukugita:1986}
M. Fukugita and T. Yanagida, ``{\em {Baryogenesis without Grand Unification}},''
  \href{http://dx.doi.org/Phys. Lett. B 174 (1986) 45}{Phys. Lett\, {\bf B\,174} (1986) 45}.

\bibitem{EWbary} 
D. E. Morrissey and M. J. Ramsey-Musolf, ``{\em {Electroweak Baryogenesis}},''
  \href{http://dx.doi.org/New J. Phys.
14, 125003 (2012).}{New\, J.\, Phys.\,{\bf 14}, 125003 (2012)}.
    
\bibitem{Buchmuller:2005eh}
Buchmuller, W. and Peccei, R.D. and Yanagida, T., ``{\em {Leptogenesis as the origin of matter }},'' 
  \href{http://dx.doi.org/10.1146/annurev.nucl.55.090704.151558}{Ann.\,Rev.\,Nucl.\,Part.\,Sci.\, {\bf 55} (2005) 311-355}.
  \href{http://arxiv.org/abs/0502169}{arXiv:0502169 [hep-ph]}.
\bibitem{Davidson:2008bu}
Davidson, Sacha and Nardi, Enrico and Nir, Yosef, ``{\em {Leptogenesis}},''
  \href{http://dx.doi.org/10.1016/j.physrep.2008.06.002}{Phys.\,Rept.\,{\bf 466} (2008) 105-177}.
  \href{http://arxiv.org/abs/0802.2962}{arXiv:0802.2962 [hep-ph]}.

\bibitem{Babu:2012vc}
Babu, K. S. and Mohapatra, R. N., ``{\em {Coupling Unification, GUT-Scale Baryogenesis and 
   Neutron-Antineutron Oscillation in SO(10)}},''
  \href{http://dx.doi.org/10.1016/j.physletb.2012.08.006}{Phys.\,Lett.\,{\bf B\,715} (2012) 328-334}.
  \href{http://arxiv.org/abs/1206.5701}{arXiv:1206.5701 [hep-ph]}.
\bibitem{Babu:2013yca}
Babu, K. S. and Bhupal Dev, P. S. and Fortes, Elaine C. F. S. and Mohapatra, R. N.",
      ``{\em {Post-Sphaleron Baryogenesis and an Upper Limit on the Neutron-Antineutron 
      Oscillation Time}},''
  \href{http://dx.doi.org/10.1103/PhysRevD.87.115019}{Phys.\,Rev.\,{\bf D\,87} (2013) 115019}.
  \href{http://arxiv.org/abs/1303.6918}{arXiv:1303.6918 [hep-ph]}.
\bibitem{Babu:2006xc}
Babu, K. S. and Mohapatra, R. N. and Nasri, S.", ``{\em {Post-Sphaleron Baryogenesis}},''
  \href{http://dx.doi.org/10.1103/PhysRevLett.97.131301}{Phys.\,Rev.\,Lett.\,{\bf 97} (2006) 131301}.
  \href{http://arxiv.org/abs/0606144}{arXiv:0606144 [hep-ph]}.
\bibitem{Mohapatra:1974gc}
R.~Mohapatra and J.~C. Pati, ``{\em {A Natural Left-Right Symmetry}},''
  \href{http://dx.doi.org/10.1103/PhysRevD.11.2558}{Phys.Rev. {\bf D\,11}, 2558 (1975)}.
\bibitem{Pati:1974yy}
J.~C. Pati and A.~Salam, ``{\em {Lepton Number as the Fourth Color}},''
  \href{http://dx.doi.org/10.1103/PhysRevD.10.275,
  10.1103/PhysRevD.11.703}{Phys. Rev. {\bf D\,10}, 275 (1974)}.
\bibitem{Senjanovic:1975rk}
G.~Senjanovic and R.~N. Mohapatra, ``{\em {Exact Left-Right Symmetry and
  Spontaneous Violation of Parity}},''
  \href{http://dx.doi.org/10.1103/PhysRevD.12.1502}{Phys. Rev. {\bf D\,12},1502 (1975)}.
\bibitem{Mohapatra:1980yp}
R.~N. Mohapatra and G.~Senjanovic, ``{\em {Neutrino Masses and Mixings in Gauge
  Models with Spontaneous Parity Violation}},''
\href{http://dx.doi.org/10.1103/PhysRevD.23.165}{Phys.Rev. {\bf D23} (1981)
  165}.

\bibitem{Mohapatra:1979ia}
R.~N. Mohapatra and G.~Senjanovic, ``{\em {Neutrino Mass and Spontaneous Parity
  Violation}},''
\href{http://dx.doi.org/10.1103/PhysRevLett.44.912}{Phys.Rev.Lett. {\bf 44}
  (1980)  912}.

\bibitem{Deshpande:1990ip}
N.~Deshpande, J.~Gunion, B.~Kayser, and F.~I. Olness, ``{\em {Left-right
  symmetric electroweak models with triplet Higgs}},''
\href{http://dx.doi.org/10.1103/PhysRevD.44.837}{Phys.Rev. {\bf D44} (1991)
  837--858}.
\bibitem{Lazarides:1980nt}
G.~Lazarides, Q.~Shafi, and C.~Wetterich, ``{\em {Proton Lifetime and Fermion
  Masses in an SO(10) Model}},''
\href{http://dx.doi.org/10.1016/0550-3213(81)90354-0}{Nucl.Phys. {\bf B181}
  (1981)  287}.
\bibitem{Dev:2013oxa}
P.~S.~ Bhupal Dev and C.~-Hun ~Lee and R.~N.~ Mohapatra, 
  ``{\em {Natural TeV-Scale Left-Right Seesaw for Neutrinos and Experimental Tests}},''
    \href{http://arxiv.org/abs/1309.0774}{{\tt arXiv:1309.0774 [hep-ph]}}.
\bibitem{Awasthi:2013ff}
R.~L. Awasthi, M.~Parida, and S.~Patra, ``{\em {Neutrino masses, dominant
  neutrinoless double beta decay, and observable lepton flavor violation in
  left-right models and SO(10) grand unification with low mass $ W_R, Z_R$
  bosons}},''
  \href{http://dx.doi.org/10.1007/JHEP08(2013)122}{JHEP\, {\bf 08} (2013) 122}.
  \href{http://arxiv.org/abs/1302.0672}{arXiv:1302.0672 [hep-ph]}.
\bibitem{typeI}
P.~Minkowski,
{Phys. Lett.} {\bf B67}, 421 (1977);
M.~Gell-Mann, P.~Ramond, and R.~Slansky (1980), print-80-0576 (CERN);
T.~Yanagida (1979), in Proceedings of the Workshop on the Baryon Number of the Universe and Unified Theories, Tsukuba, Japan, 13-14 Feb 1979;
R. N.~Mohapatra and G.~Senjanovic,
{Phys. Rev. Lett} {\bf 44}, 912 (1980);
J.~Schechter and J. W. F.~Valle,
{Phys. Rev.} {\bf D22}, 2227 (1980).
    
\bibitem{inv}
R.~N.~Mohapatra,
Phys.\ Rev.\ Lett.\  {\bf 56}, 561 (1986);
 R.~N.~Mohapatra, J.~W.~F.~Valle,
 Phys.\ Rev.\  {\bf D\,34}, 1642 (1986).
\bibitem{Bdev-non} P.~S.~B.~Dev, R.~N.~Mohapatra,
                   Phys.\ Rev.\  {\bf D\,81}, 013001 (2010); arXiv:0910.3924 [hep-ph].
\bibitem{pdg} K. Nakamura {\em et al.} (Particle Data Group), J. Phys. {\bf G\,37}, 075021 (2010); 
              C. Amsler {\em et al.} (Particle Data Group), Phys. Lett. {\bf B\,667}, 1 (2008).
\bibitem{ATLAS:2012ak}
{\bf ATLAS Collaboration}, G.~Aad {\em et al.}, ``{\em {Search for heavy
  neutrinos and right-handed $W$ bosons in events with two leptons and jets in
  $pp$ collisions at $\sqrt{s}=7$ TeV with the ATLAS detector}},''
  \href{http://dx.doi.org/10.1140/epjc/s10052-012-2056-4}{Eur. Phys. J. {\bf
  C72} (2012)  2056},
\href{http://arxiv.org/abs/1203.5420}{{\tt arXiv:1203.5420 [hep-ex]}}.

\bibitem{CMS:2012zv}
{\bf CMS Collaboration}, S.~Chatrchyan {\em et al.}, ``{\em {Search for heavy
  neutrinos and W[R] bosons with right-handed couplings in a left-right
  symmetric model in pp collisions at sqrt(s) = 7 TeV}},''
  \href{http://dx.doi.org/10.1103/PhysRevLett.109.261802}{Phys.Rev.Lett. {\bf
  109} (2012)  261802},
\href{http://arxiv.org/abs/1210.2402}{{\tt arXiv:1210.2402 [hep-ex]}}.
\bibitem{Babu:1992ia}
K.~S.~ Babu and R.~N.~ Mohapatra, 
  ``{\em {Predictive neutrino spectrum in minimal $SO(10)$ grand unification }},''
    \href{http://dx.doi.org/10.1103/PhysRevLett.70.2845}{Phys.\,Rev.\,Lett.\, {\bf 70} (1993) 2845}.
    \href{http://arxiv.org/abs/9209215}{{\tt arXiv:9209215 [hep-ph]}}.

\bibitem{Bertolini:2013vta}
S.~ Bertolini, L.~ Di Luzio and M.~Malinsky,  
  ``{\em {Light color octet scalars in the minimal $SO(10)$ grand unification }},''
    \href{http://dx.doi.org/ }{Phys.\,Rev. {\bf D\,87} (2013) 085020}.
      \href{http://arxiv.org/abs/1302.3401}{{\tt arXiv:1302.3401 [hep-ph]}}.
      
\bibitem{Ibanez:1984ni}
Luis E.\,Ibanez and C.\,Munoz, 
``{\em {Enhancement Factors for Supersymmetric Proton Decay in the Wess-Zumino Gauge }},''
  \href{http://dx.doi.org/10.1016/0550-3213(84)90439-5 }{Nucl. Phys. {\bf B\,245} (1984) 425}.

\bibitem{Buras:1977yy}
A.J.\,Buras, John R.\,Ellis, M.K.\,Gaillard and Dimitri V.\,Nanopoulos, 
``{\em {Aspects of the Grand Unification of Strong, Weak and Electromagnetic Interactions }},''
  \href{http://dx.doi.org/10.1016/0550-3213(78)90214-6 }{Nucl. Phys. {\bf B\,135} (1978) 66-92}.

\bibitem{BhupalDev:2010he}
P.S.\, Bhupal Dev and R.N.\, Mohapatra, 
``{\em {Electroweak Symmetry Breaking and Proton Decay in $SO(10)$
                        SUSY-GUT with TeV $W_R$ }},''
\href{http://dx.doi.org/10.1103/PhysRevD.82.035014}{Phys.\,Rev. {\bf D\,82} (2010) 035014}.
\href{http://arxiv.org/abs/1003.6102}{{\tt arXiv:1003.6102 [hep-ph]}}.
%


%






\bibitem{Nishino:2012ipa}
{\bf Super-Kamiokande Collaboration}, H.~ Nishino {\em et al.}
  ``{\em {Search for Nucleon Decay into Charged Anti-lepton plus Meson in Super-Kamiokande I and II }},''
    \href{http://dx.doi.org/10.1103/PhysRevD.85.112001}{Phys.\,Rev. {\bf D\,85} (2012) 112001}.
      \href{http://arxiv.org/abs/1203.4030}{{\tt arXiv:1203.4030 [hep-ph]}}.

\bibitem{babuetal} 
K. S. Babu {\em et al.}, ``{\em  {Proton Decay, presented at the workshop on Fundamental 
      Physics at the Intensity Frontier, Rockville, Maryland, Nov.\,20 - Dec.\,2, 2011}}'',
      \href{http://arxiv.org/abs/1205.2671}{\tt arXiv:1205.2671 [hep-ex]}.
\bibitem{hyperk}
K. Abe et al., (2011), 
     ``{\em  {Letter of Intent: The Hyper-Kamiokande Experiment --- Detector Design and Physics Potential}}'',     
     \href{http://arxiv.org/abs/1109.3262}{\tt arXiv:1109.3262 [hep-ex]}.

\end{thebibliography}
\end{document}